\RequirePackage{silence}

\documentclass[11pt]{article}
\usepackage[in]{fullpage}
\usepackage{parskip}

\usepackage{amsmath,amssymb,amsthm,mathtools}

\usepackage{Levi}
\usepackage{alglib-tomer}
\usepackage{xspace}
\usepackage{svg}
\usepackage{colortbl}
\usepackage{tocloft}
\usepackage{makecell}
\usepackage{booktabs}
\definecolor{bestcolor}{RGB}{220,240,255} 
\usepackage[capitalize]{cleveref}

\makeatletter
\newtheorem*{rep@theorem}{\rep@title}
\newcommand{\newreptheorem}[2]{%
	\newenvironment{rep#1}[1]{%
		\def\rep@title{#2 \ref{##1}}%
		\begin{rep@theorem}}%
		{\end{rep@theorem}}}
\newcommand{\ignore}[1]{}        
\makeatother
\newreptheorem{theorem}{Theorem}
\newreptheorem{example}{Example}
\newreptheorem{claim}{Claim}

\newcommand\abs[1]{\left|{#1}\right|}
\newcommand\floor[1]{\left\lfloor{#1}\right\rfloor}
\newcommand\ceil[1]{\left\lceil{#1}\right\rceil}

\renewcommand\eps{\varepsilon}

\newcommand{\oracleIS}{\textsf{IS}\xspace}
\newcommand{\oracleDeg}{\textsf{Deg}\xspace}
\newcommand{\oracleN}{\textsf{Neigh}\xspace}
\newcommand{\boldm}{\boldsymbol{m}}

\newcommand\mbar{\texorpdfstring{\ensuremath{\bar{m}}}{mbar}}

\newcommand\procnameZenumerateZedges{\textsf{Enumerate-Edges}\xspace}

\newcommand\procnameZestimateZedgesZiteration{\textsf{Estimate-Edges-Iteration}}
\newcommand\procnameZestimateZedgesZiterationHREF{\hyperref[alg:e-e-iteration]{\procnameZestimateZedgesZiteration}\xspace}

\newcommand\procnameZestimateZedgesZbounded{\textsf{Estimate-Edges-Bounded}}
\newcommand\procnameZestimateZedgesZboundedHREF{\hyperref[alg:e-e-bounded]{\procnameZestimateZedgesZbounded}\xspace}

\newcommand\procnameZestimateZedges{\textsf{Estimate-Edges}}
\newcommand\procnameZestimateZedgesHREF{\hyperref[alg:e-e]{\procnameZestimateZedges}\xspace}

\newcommand\procnameZestimateZedgesZadviceZhigh{\textsf{Estimate-Edges-Advice-High}}
\newcommand\procnameZestimateZedgesZadviceZhighHREF{\hyperref[alg:e-e-advice-high]{\procnameZestimateZedgesZadviceZhigh}\xspace}

\newcommand\procnameZestimateZedgesZadviceZlow{\textsf{Estimate-Edges-Advice-Low}}
\newcommand\procnameZestimateZedgesZadviceZlowHREF{\hyperref[alg:e-e-advice-low]{\procnameZestimateZedgesZadviceZlow}\xspace}

\newcommand\procnameZsampleZllZedgesZsparse{\textsf{Sample-LL-Edges-Sparse}}
\newcommand\procnameZsampleZllZedgesZsparseHREF{\hyperref[alg:e-ll-sparse]{\procnameZsampleZllZedgesZsparse}\xspace}

\newcommand\procnameZestimateZllZedgesZadvice{\textsf{Estimate-LL-Edges-Advice}}
\newcommand\procnameZestimateZllZedgesZadviceHREF{\hyperref[alg:e-ll-advice]{\procnameZestimateZllZedgesZadvice}\xspace}

\newcommand\procnameZestimateZllZedgesZadviceZbounded{\textsf{Estimate-LL-Edges-Advice-Bounded}}

\newcommand\procnameZestimateZLoneHZedgesZadvice{\textsf{Estimate-L${}_1$H-Edges-Advice}}
\newcommand\procnameZestimateZLoneHZedgesZadviceHREF{\hyperref[alg:e-l1h-advice]{\procnameZestimateZLoneHZedgesZadvice}\xspace}

\newcommand\procnameZguardZquantity{\textsf{Guard-Quantity}}
\newcommand\procnameZguardZquantityHREF{\hyperref[fig:alg:small-mbar-guard-APX-CASE-QUANTITY]{\procnameZguardZquantity}\xspace}

\newcommand\procnameZguardZquality{\textsf{Guard-Quality}}
\newcommand\procnameZguardZqualityHREF{\hyperref[fig:alg:small-mbar-guard-APX-CASE-QUALITY]{\procnameZguardZquality}\xspace}

\newcommand\procnameZsmallZmbarZguard{\textsf{Small-$\bar{m}$-Guard}}
\newcommand\procnameZsmallZmbarZguardHREF{\hyperref[fig:alg:small-mbar-guard]{\procnameZsmallZmbarZguard}\xspace}

\title{When Local and Non-Local Meet: Quadratic Improvement for Edge Estimation with Independent Set Queries}
\author{ Tomer Adar\thanks{Technion - Israel Institute of Technology, Israel. Email: \href{mailto: tomer-adar@campus.technion.ac.il }{tomer-adar@campus.technion.ac.il}.}
\and Yahel Hotam\thanks{University of Haifa, Israel. Email: \href{mailto: yahelhotam@gmail.com }{yahelhotam@gmail.com}.}
\and Amit Levi\thanks{University of Haifa, Israel. Email: \href{mailto: alevi@cs.haifa.ac.il}{alevi@cs.haifa.ac.il}.} 
}
\begin{document}
\maketitle

\begin{abstract}
    We study the problem of estimating the number of edges in an unknown graph. We consider a hybrid model in which an algorithm may issue independent set, degree, and neighbor queries. We show that this model admits strictly more efficient edge estimation than either access type alone. Specifically, we give a randomized algorithm that outputs a $(1\pm\eps)$-approximation of the number of edges using $
        O\left(\min\left(\sqrt{m}, \sqrt{\frac{n}{\sqrt{m}}}\right)\cdot\frac{\log n}{\eps^{5/2}}\right)
    $
    queries, and prove a nearly matching lower bound. 
    In contrast, prior work shows that in the local query model (Goldreich and Ron, \textit{Random Structures \& Algorithms} 2008) and in the independent set query model (Beame \emph{et al.} ITCS 2018, Chen \emph{et al.} SODA 2020), edge estimation requires $\widetilde{\Theta}(n/\sqrt{m})$ queries in the same parameter regimes. Our results therefore yield a quadratic improvement in the hybrid model, and no asymptotically better improvement is possible. 
\end{abstract}
\thispagestyle{empty}

\newpage
\thispagestyle{empty}
\setcounter{tocdepth}{2}
\setlength{\cftparskip}{5pt}
\tableofcontents
\thispagestyle{empty}
\newpage
\setcounter{page}{1}

\renewcommand{\arraystretch}{1.8} 
\setlength{\tabcolsep}{15pt}

\section{Introduction}
Estimating the number of edges in an unknown graph is a central problem in sublinear algorithms. Formally, given query access to a graph $G=(V,E)$ on $n$ vertices, the goal is to approximate $m = |E|$ within a $(1\pm\eps)$-multiplicative factor while making as few queries as possible. The problem  captures the inherent tension between accuracy and information constraints and has been extensively studied (see e.g.,~\cite{F06,GR08,ER18,ERS19,AKK19,LPRV21,TT22,BT24,BCM25})

A key theme in this line of work is that the achievable query complexity depends critically on the type of access allowed to the graph. Different query models expose fundamentally different kinds of information, leading to qualitatively different algorithmic techniques.

Until recently, most work in the sublinear algorithms community focused on \emph{local access models}, where queries are restricted to a vertex and its neighborhood. Typical examples include degree queries, which return a vertex's degree, and neighbor queries, which return a specified neighbor. Such queries provide fine-grained, precise information about the graph.

Beame, Har-Peled, Ramamoorthy, Rashtchian, and Sinha~\cite{BHRRS18} introduced a markedly different access model, the \emph{independent-set (IS) oracle}. In this model, an algorithm may query any subset $S\subseteq V$, and the oracle responds with a single bit indicating whether $S$ induces an edge-free subgraph. Unlike local queries, IS queries do not reveal the identity or location of edges. Instead, they provide a coarse \emph{non-local} information about the presence or absence of edges within a queried set. This model can be viewed as a form of combinatorial group testing, where a single query can simultaneously probe a large portion of the graph. As such, IS queries expose structural information that is fundamentally inaccessible through purely local inspection.

\cite{BHRRS18}~first analyzed edge estimation in the IS model and obtained an upper bound of $\widetilde O(n^{2/3})$ queries. Chen, Levi, and Waingarten~\cite{CLW20} refined this analysis and established nearly optimal bounds: $(1\pm\eps)$-approximate edge estimation can be performed using
\[
\widetilde{O}\left(\min\left(\sqrt{m}, \frac{n}{\sqrt{m}}\right)\right)
\]
IS queries, and show that this dependence on $n$ and $m$ is optimal up to polylogarithmic factors.

Local queries and IS queries exhibit inherently different strengths. Local queries provide detailed information about individual vertices and their neighborhoods, whereas IS queries give coarse binary feedback over subsets. When considering the local model, (nearly) tight bounds of $\widetilde \Theta(n/\sqrt{m})$ queries for edge estimation are known~\cite{GR08,F06}. Therefore, there exist settings in which both oracles require the same asymptotic number of queries, but for entirely different reasons: local queries are limited by the need to sample enough edges from sparse regions, while IS queries are limited by the coarse granularity of global binary feedback. This illustrates that even when the query complexities coincide, the information revealed by each oracle is \emph{fundamentally different}, reflecting their complementary powers.

This contrast naturally motivates the study of \emph{hybrid} access models, in which an algorithm is allowed to use both local queries and IS queries. Such a model offers the potential to combine non-local aggregation with local precision: local queries can be used to identify informative regions of the graph or reveal degrees, whereas IS queries can efficiently summarize the presence of edges across large subsets. Understanding the power and limitations of this hybrid model is essential for clarifying how different forms of partial information interact, and for determining whether combining query types can lead to provable improvements over either model alone.

\subsection{Our Results}
In this work, we provide a tight characterization of this hybrid model.

\begin{theorem}[Upper Bound]
Consider the query model consisting of IS, degree, and neighbor queries to an unknown graph $G=(V,E)$ with $|V|=n$ and $|E|=m$.
For every $0 < \eps < 1$, there exists a randomized algorithm that uses
\[
O\left(\min\left(\sqrt{m}, \sqrt{\frac{n}{\sqrt{m}}}\right)\cdot\frac{\log n}{\eps^{5/2}}\right)
\]
queries in expectation and outputs a $(1\pm\eps)$-approximation of $m$ with probability at least $2/3$.
\end{theorem}
 
\begin{theorem}[Lower Bound]
For every $\eps \in (0,1)$, in the query model allowing independent set, degree, and neighbor queries, any randomized algorithm that estimates the number of edges $m$ of an $n$-vertex graph to within a $(1\pm\eps)$ factor with success probability at least $2/3$ requires
\[
\Omega\left(\min\left(\sqrt{m}, \sqrt{\frac{n}{\eps\sqrt{m}}}\right)\right)
\]
queries.
\end{theorem}

In particular, in regimes where each individual model would require $\widetilde\Theta(n/\sqrt{m})$ queries, the combination yields a \emph{quadratic improvement}, whereas in other regimes it matches the best of the two models. These results demonstrate that IS and local queries are synergistic yet complementary, and that the hybrid model achieves the maximum possible improvement for the task of edge estimation. 

\subsection{Technical overview}
\subsubsection{Upper-Bound}

Our algorithm is advice-based: it receives a guess $\bar m$ for the number of edges $m$, but treats this guess as untrusted. For each guess, the algorithm must either produce a $(1\pm\eps)$-approximation to $m$, or determine that $\bar m$ is not within the correct scale. The goal is to obtain query complexity
\[
\widetilde{O}\left(\min\left(\sqrt{m}, \sqrt{n / \sqrt{m}}\right)\right).
\]

To achieve this goal, we interleave two exponentially progressing sequences of guesses. The upper sequence starts from $n^2$ and decreases geometrically, taking $\bar m = 2^{-\ell}n^2$ at iteration $\ell$. The lower sequence increases geometrically, taking $\bar m = 2^{\ell/2}$. The different rates are chosen so that the cost of handling a guess from either sequence is comparable within the same iteration.

For a fixed guess $\bar m$, the algorithm partitions vertices according to a degree threshold $k$, classifying vertices as low-degree or high-degree. This induces a decomposition of the edge set into edges between two low-degree vertices ($m_{LL}$), edges between low- and high-degree vertices ($m_{LH}$), and edges between two high-degree vertices ($m_{HH}$). These components are handled separately, exploiting the fact that high-degree vertices are rare once $k$ is chosen appropriately.

The main component is estimating $m_{LL}$. We adapt the edge-sampling approach of \cite{BHRRS18}, which samples an induced subgraph by including each vertex independently with probability $1/\sqrt{\bar m}$. If all vertices were low-degree, then this estimator would have low variance. We use the degree oracle to identify edges incident to high-degree vertices and exclude them from counting to keep the variance low while preserving unbiasedness.

Edges between high-degree vertices have small contribution overall. For an appropriate choice of $k$, the number of such edges is $O(\eps m)$, which can be absorbed into the allowed additive error. For low-high edges, we further refine the low-degree vertices by introducing a second threshold $k'$, defining very-low-degree vertices, and only count the edges between them to high-degree vertices. With this refinement, for an appropriate choice of $k$, the number of excluded edges is $O(\eps m)$, allowing us to only consider edges between very-low-degree vertices and high-degree vertices.

The remaining quantity, denoted $m_{L_1H}$ (between very-low-degree vertices and high-degree vertices), is estimated via a weighted sampling procedure. We sample a vertex, then sample one of its neighbors, and count the edge with weight proportional to the degree of the sampled vertex whenever it is very-low-degree and its neighbor is high-degree. This yields an unbiased estimator whose variance is controlled using $O(n/(\eps^3 k))$ samples.

We analyze the algorithm separately for the two guess sequences. We first consider the upper guess sequence, for which $\bar m \ge m$ holds throughout until a correct-scale guess is reached. In this regime, we set 
\[
k = \Theta\left(\sqrt{n\sqrt{\bar m}/\eps}\right),
\]
except in the very dense case $\bar m = \Omega(\eps n^2)$, where a separate optimization applies. With this choice, the combined cost of estimating $m_{LL}$ and $m_{L_1H}$ is
\[
O\left(\frac{\log n}{\eps^{5/2}} \cdot \sqrt{\frac{n}{\sqrt{\bar m}}}\right).
\]
When $\bar m = \Theta(m)$, the algorithm outputs $\tilde m = (1\pm O(\eps))m$ with high probability. More generally, when $\bar m$ is larger than the true value, the output is noticeably smaller than $\bar m$, and in particular bounded by $m + \eps \bar m \ll \bar m$ when $\bar m \gg m$. This gap allows the algorithm to reliably detect overly large guesses and continue to smaller ones.

For the lower guess sequence, the algorithm cannot assume that $\bar m \ge m$. Instead, it sets the degree threshold to $k=\min(\bar m,n-1)$ and attempts to verify that no vertex exceeds this degree. If this verification succeeds, then all vertices are low-degree and $m_{LL}=m$, so estimating $m_{LL}$ alone suffices. If the guess $\bar m$ is too small, then either $m_{LL} > 2 \bar m$, or there exist $\Omega(m)$ edges adjacent to vertices of degree larger than $\bar m$ (in particular, there exists a vertex of degree larger than $\bar m$). These cases are detected by running a complexity-safe version of the $m_{LL}$ estimator, which aborts if it exceeds the cost expected when assuming that $\bar m \ge m$, and by searching for a high-degree vertex. The resulting cost per lower-sequence iteration is
\[
O\left(\frac{\log n}{\eps^2}\sqrt{\bar m}\right).
\]

Standard amplification could be applied to each iteration to ensure correctness with high probability. However, a more refined analysis shows that such amplification is unnecessary. The failure probability of an upper-sequence iteration is $O(m/\bar m)$, while that of a lower-sequence iteration is $O(\bar m/m)$. Since both guess sequences are exponential, the total failure probability accumulated over all iterations preceding the one with $\bar m \approx m$ remains bounded by a constant. The iteration for which $\bar m = \Theta(m)$ therefore succeeds with constant probability.

Finally, while the above discussion yields the desired query complexity with high probability, controlling the expected complexity requires care due to the possibility that ``the'' correct guess fails, and then further high-sequence guesses are smaller than $m$ and low-sequence guesses are greater than $m$, making the rest of the run incompatible with the correctness analysis (even though complexity analysis still holds). To address this, we execute the guesses using an iterative deepening schedule. We first consider iteration $0$, then iterations $(0,1)$, then $(0,1,2)$, and so on. This scheduling ensures that the tail of the running-time distribution decays sufficiently fast, so that the expected query complexity matches the high-probability bound asymptotically.

\subsubsection{Lower bound}
We show that any deterministic algorithm with access to degree, neighbor and independent-set oracles cannot distinguish between two distributions over graph whose degrees are separated by a $(1+\Omega(\eps))$-factor, unless it makes $\Omega\left(\min\left(\sqrt{m}, \sqrt{n / \eps\sqrt{m}}\right)\right)$ queries in expectation (note that the input distribution is the source of randomness). By Yao's principle \cite{Yao77}, this lower bound also applies to probabilistic algorithms with the same expected query complexity.

\paragraph{Canonical graph construction} To define the canonical graph $G_{n,d,\eps}$, let $d_0(n,\epsilon) \eqdef  1/(n^{1/3}\epsilon^{2/3})$. For the simplicity of the overview, we focus on the case where $d \ge d_0(n,\eps)$, whose lower bound is $\Omega(\sqrt{n / \eps \sqrt{m}})$. The input graph consists of disjoint vertex sets $K$, $L$, $H$, and $A$ whose sizes are:
\[ \begin{aligned}
    & n_k = \floor{\sqrt{2 n d}}, 
    & n_h = \ceil{\eps^{1/2} n^{1/4} d^{3/4}}, \\
    & n_\ell = \ceil{\eps^{1/2} n^{3/4} d^{1/4}}, 
    & n_a = n - (n_k + n_h + n_\ell),
\end{aligned} \]
For the graph edges: the set $K$ forms a clique, the sets $L$ and $H$ are the sides of a biclique and $A$ consists of isolated vertices. The $K$-clique contributes $\Omega(m)$ edges (but not more than $m$), and the $L$-$H$ graph contributes $\Theta(\eps m)$ edges. Therefore, the number of edges is $\Theta(m)$ and the average degree is $\Theta(d)$.

\paragraph{Graph distributions} In the first distribution we permute the vertices of $G_{n,d,0}$, that is, the graph only consists of a clique but not a biclique, and it has at most $m$ edges. In the second distribution we permute the vertices of $G_{n,d,\eps}$, which has at least $\eps m$ additional edges.

\paragraph{Intuition} As long as the algorithm does not query a vertex in the biclique or find a biclique-edge, the input distributions are indistinguishable. For local queries, the probability of hitting the biclique scales with the size of $L$. For independent-set queries, a query must include at least one vertex from $H$ to detect an edge in the biclique. Large IS-queries are limited by the clique size $\abs{K}$, since including two vertices from $K$ makes the queried set not-independent regardless of whether or not it has a biclique-edge.

A more careful argument shows that for degree and neighbor queries, the probability of sampling a biclique vertex is $O((n_\ell + n_h)/n) = O(n_\ell/n)$. Hence, $\Omega\left(\frac{n}{n_\ell}\right) = \Omega\left(\sqrt{\frac{n}{\eps \sqrt{m}}}\right)$ queries are needed to hit a biclique vertex with $\Omega(1)$ probability.

For IS-queries, the clique $K$ limits the effective query size to $O(n/n_k) = O(\sqrt{n/d})$, and the probability of hitting $H$ using such query is $O\left(\frac{n_h}{n}\cdot \sqrt{\frac{n}{d}}\right)=O\left(\frac{n_h}{\sqrt{nd}}\right)$. This probability leads to the same lower bound as before. Overall, as $d= \Theta(m/n)$, distinguishing the two distributions with probability $\Omega(1)$ requires $\Omega\left(\sqrt{\frac{n}{\eps \sqrt{m}}}\right)$ queries, which establishes the desired lower bound.

For $d < d_0(n,\epsilon)$, we observe that the definition of $n_h$ is fixed to $1$, and therefore, we can reduce $n_\ell$ to $\ceil{\eps n d}$. Besides the alternative value of $n_\ell$, the construction is the same. Analyzing this small-$d$ case uses similar ideas and results in an $\Omega(\sqrt{m})$ lower bound.

\subsection{Related work}
A generalization of independent set queries appears in a broader line of work investigating the relationship between decision and counting problems~\cite{Sto83,Sto85,RT16, DL21,DLM22,BBGM24}. Independent set queries also have deep connections to group testing, where one tests large collections of items to detect the presence of rare events~\cite{Dor43,S85,CS90}. Independent set queries also have close connections to learning theory. In particular, a classical line of work on {learning hidden graphs} studies \emph{edge-detecting queries}, which are equivalent to independent set queries~\cite{AC08,ACW06,AB19}.

Several works have explored combinatorial or global query primitives beyond purely local access. Beyond standard IS queries, Beame \emph{et al.}~\cite{BHRRS18} introduced \emph{bipartite independent set (BIS) queries}, which ask whether there exists at least one edge between two given vertex subsets, and used them to obtain tight sublinear bounds for edge count estimation. Subsequently,~\cite{AMM22} designed \emph{non-adaptive} algorithms for edge counting and sampling in the BIS model. The BIS model was later extended to more general settings (see e.g.,~\cite{BBGM19,BGMP19,BBGM21}).

Perhaps the work most closely related to ours is that of Beretta, Chakrabarty, and Seshadhri~\cite{BCS26SODA}, which studies average-degree estimation under a query model combining random-edge sampling with structural queries (e.g., pair queries and full neighborhood access). They also discuss a structural oracle that, given a set $S \subseteq V$, returns the number of edges induced by $S$ at a cost proportional to $|S|$. These structural queries provide precise local or aggregate information but follow a different cost model than IS queries. In particular, it is possible to show that IS oracle can simulate their structural oracle using (almost) linear number of IS queries.

\section{Preliminaries}
Given a positive integer $n$, we denote $[n] = \{1,...,n\}$. For a set $A$, we denote the power set of $A$ by $2^A$. All graphs $G=(V, E)$ considered in this paper are undirected and simple (meaning that there are no parallel edges or loops), with $V=[n]$ as their vertex set (we often just write $G=([n], E)$). We also let $m=|E|$ be the number of edges in the graph and $d=2m/n$ be the average degree in the graph. We use $\deg(v)$ to denote the degree of a vertex $v\in [n]$.  We use boldface letters (such as $\bX$) to denote random variables. We use the notation $\tilde{x}=(1\pm \eps)x$ to denote that $(1-\eps)x\le \tilde x\le (1+\eps)x$.

An algorithm is allowed to perform three types of queries to access the (otherwise hidden) edge set of the graph $E$, as defined by the following oracles:

\begin{definition}[Degree oracle]
    Given an undirected graph $G=([n], E)$, its \emph{degree oracle} is a map $\oracleDeg: [n] \rightarrow [n-1]\cup\{0\}$, which for a vertex $u\in [n]$ returns the degree $\deg(u)$.
\end{definition}

\begin{definition}[Random Neighbor oracle]
    Given an undirected graph $G=([n], E)$, its \emph{random neighbor oracle} is a map $\oracleN: [n] \rightarrow [n]$, which for a vertex $u\in [n]$ returns a uniformly random neighbor of $u$. If $u$ has no neighbors, it returns a special character to signify it.
\end{definition}

\begin{definition}[Independent set oracle]
    Given an undirected graph $G=([n], E)$, its \emph{independent set oracle} is a boolean map $\oracleIS: 2^{[n]} \rightarrow\{0,1\} $, which for a subset of vertices $U\subseteq[n]$ returns 1 if and only if $U$ is an independent set of $G$ (i.e., the subgraph induced on $U$ contains no edges).
\end{definition}

Some of our algorithms will include the assumption that $m\geq 1$, note that we can distinguish this from the case that $m=0$ using a single query to \oracleIS.

\subsection{Classifying vertices and edges}
We classify the vertices in $V$ into three disjoint sets by their degrees, and use those sets to classify the edges in the graph:
\begin{definition} \label{def:vertex-classification}
    Given a graph $G=(V, E)$, let $k'$ and $k$ be two degree thresholds (not necessarily integers) for which $k \ge k' \ge 0$ (where $n = \abs{V}$). Let $V_{L_1}, V_{L_2}, V_H$ be a partition of $[n]$ into disjoint sets such that:
    \[  V_{L_1} =\left\{ u\in V : \deg (u) \le k' \right\},\quad
        V_{L_2} =\left\{ u\in V : k' < \deg (u) \le k \right\},\quad
        V_H =\left\{ u\in V : \deg (u) > k \right\}.
    \]
    Also, let $V_L = V_{L_1} \cup V_{L_2} = \{ u \in V : \deg (u) \le k\}$.

    For two labels $\sigma_1,\sigma_2 \in \{ L_1, L_2, L, H \}$, we use $E_{\sigma_1 \sigma_2}$ to denote the set of edges between $V_{\sigma_1}$ and $V_{\sigma_2}$ (note that $E_{\sigma_1 \sigma_2}$ and $E_{\sigma_2 \sigma_1}$ are identical). Also, we let $m_{\sigma_1 \sigma_2} = \abs{E_{\sigma_1 \sigma_2}}$ be the number of these edges.
\end{definition}

At top-level, we ``guess'' (formally, search for) an advice $\bar{m}$, hopefully just slightly greater than the correct answer $m$. Based on $\bar{m}$ we define the threshold degrees $k$ and $k'$, and use different logic to estimate the number of edges between every pair of subsets. For some pairs the estimation is zero, since the choice of $k,k'$ means that the number of relevant edges is no more than linear in the allowed additive error and thus are negligible.

For $\bar{m} > 0$ and $0 < \eps < 1$, let  $k(\bar{m},\eps) \eqdef \sqrt{\frac{2n\sqrt{\bar m}}{\eps}}$ and $k' = \min(k, \bar{m} / \eps k)$, and recall the sets of Definition \ref{def:vertex-classification}. Clearly, both $k$ and $\bar{m} / \eps k$ are monotone increasing in $\bar{m}$ (when $n$ and $\eps$ are considered fixed). Therefore, the set $V_H(\bar{m},\eps)$ and the union set $V_{L_2}(\bar{m},\eps) \cup V_H(\bar{m},\eps)$ are monotone non-increasing in $\bar m$. Note that when $k'=k$, we have $|V_{L_2}|=0$ and thus $V_L = V_{L_1}$.

\begin{lemma} \label{lm:just-few-edges-above-kprime-to-high}
    Let $k(\bar{m},\eps) = \sqrt{2 n \sqrt{\bar{m}} / \eps}$. If $\bar{m} \ge \frac{1}{4}m$, then the number of edges between $V_{L_2} \cup V_H$ and $V_H$ is at most $64 \eps \bar{m}$.
\end{lemma}
\begin{proof}
    There are two cases: $k' = \mbar / \eps k$ and $k' = k$.
    If $k' = \mbar / \eps k$, then the number of vertices in $V_{L_2} \cup V_H$ is at most $\frac{2m}{\bar{m} / (\eps k)} = \frac{2 \eps m k}{\bar{m}}$. The number of vertices in $V_H$ is at most $2 m / k$. Hence, the number of edges between $V_{L_2} \cup V_H$ and $V_H$ is at most 
    \[  \frac{2 \eps m k}{\bar{m}} \cdot \frac{2 m}{k}
        = 4 \eps m \cdot \frac{m}{\bar{m}}
        = 4\eps \left(\frac{m}{\bar{m}}\right)^2 \bar{m}
        \le 64 \eps \mbar.
    \]

    If $k' = k$, then the number of vertices in $V_{L_2} \cup V_H$ (which is actually $V_H$) is at most $\frac{2m}{k}$, and therefore,
    \begin{eqnarray*}
        \left(\frac{2m}{k}\right)^2
        = \frac{4m^2}{k^2}
        = \frac{4m^2}{2n \sqrt{\mbar} / \eps}
        = 2\eps \cdot \frac{m^2}{n \sqrt{\mbar}}
        \le 2\eps \cdot \frac{m^2}{\sqrt{m} \cdot \sqrt{m/4}}
        = 4\eps m
        \le 16\eps \mbar.
    \end{eqnarray*}
\end{proof}

\begin{lemma} \label{lm:mll-ml1h-cover-most}
    If $\bar{m} \ge \frac{1}{4} m$ and $k(\bar{m},\eps) = \sqrt{2 n \sqrt{\bar{m}} / \eps}$, then $m_{LL} + m_{L_1 H} \ge m - 64 \eps\bar m$.
\end{lemma}
\begin{proof}
    Partition the edges as $m = m_{LL} + m_{L_1 H} + m_{L_2 H} + m_{HH}$. By Lemma~\ref{lm:just-few-edges-above-kprime-to-high}, $m_{L_2 H} + m_{HH} \le 64 \eps \bar{m}$. Subtracting these edges gives $m_{LL} + m_{L_1 H} = m - (m_{L_2 H} + m_{HH}) \ge m - 64 \eps \bar{m}$.
\end{proof}

\section{Upper Bound}

\newcommand\reject{\textsc{reject}\xspace}

\newcommand\mbig{{\ensuremath{\bar{m}_\mathrm{big}}}}
\newcommand\msmall{{\ensuremath{\bar{m}_\mathrm{small}}}}

\newcommand\ellbig{{\ensuremath{\ell_\mathrm{big}}}}
\newcommand\ellsmall{{\ensuremath{\ell_\mathrm{small}}}}

\newcommand\tildembig{{\ensuremath{\tilde{\bm}_\mathrm{big}}}}
\newcommand\tildemsmall{{\ensuremath{\tilde{\bm}_\mathrm{small}}}}

In this section, we prove an upper bound on the number of queries an algorithm with access to \oracleDeg, \oracleN, \oracleIS needs to make in order to achieve a $1\pm\eps$ estimation of the number of edges in a graph. Specifically, we prove the following:

\begin{theorem} \label{untrusted--thm:ubnd}
    There exists an algorithm $\procnameZestimateZedgesHREF(n, \eps, G)$ that takes as input integer $n\geq 1$, parameter $\eps\in(0,1]$, and access to the oracles \oracleDeg, \oracleN, \oracleIS of a graph $G=([n], E)$ with $m=|E|$. \procnameZestimateZedgesHREF makes an expected number of $O\left(\frac{\log n}{\eps^{5/2}} \min\left(\sqrt{\frac{n}{\sqrt{m}}}, \sqrt{m} \right)\right)$ queries to the oracles and outputs a number $\tilde \boldm$ such that with  probability at least $2/3$ satisfies $(1-\eps)m \leq \tilde \boldm \leq (1+\eps)m$.
\end{theorem}
We prove Theorem \ref{untrusted--thm:ubnd} through Lemma \ref{lem:ubnd}, which is logically equivalent. We use the parameters and notation described in \cref{tab:parameters} throughout the paper.

\begin{table}[ht]
\centering
\renewcommand{\arraystretch}{1.15}
\begin{tabular}{|c|p{11cm}|}
\hline
\textbf{Symbol} & \textbf{Description} \\
\hline
$G=(V,E)$
& The input graph. \\

$n$
& Number of vertices in the graph ($n = |V|$). \\

$m$
& Number of edges in the graph ($m = |E|$). \\

$\bar{m}$
& Untrusted advice for the number of edges $m$, where $\bar{m} > 0$. \\

$\tilde{\bm}$
& Random variable output of an estimation procedure.\\

$k$
& Degree threshold: vertices of degree at most $k$ are \emph{low-degree}, and
vertices of degree greater than $k$ are \emph{high-degree}, with
$k \ge 0$. \\

$k'$
& Secondary degree threshold, satisfying $0 \le k' \le k$. \\
\hline
\end{tabular}
\caption{Notation and parameters used throughout this section.}
\label{tab:parameters}
\end{table}

\subsection{Estimating $m_{LL}$ (with advice)}

We restate the following lemma from~\cite{BHRRS18}:

\begin{lemma}[\cite{BHRRS18}] \label{lm:enumerate-edges}
    There exists  a deterministic procedure \procnameZenumerateZedges whose input is a graph $G = (V,E)$ accessible through the independent-set oracle and a vertex set $A \subseteq V$, and its output is a list of all edges in the subgraph induced on $A$'s vertices. The query complexity of this procedure is $O(1 + \abs{E_A} \log n )$, where $n = \abs{V}$ is the number of vertices and $E_A$ is the set of edges in the subgraph induced on $A$'s vertices.
\end{lemma}
Procedure \procnameZsampleZllZedgesZsparseHREF, implemented in Algorithm  \ref{alg:e-ll-sparse}, is an unbiased estimator for $\frac{m_{LL}}{\bar{m}}$. It draws a set $\bS \subseteq V$ that every vertex belongs to with probability $1/\sqrt{\bar{m}}$, and then enumerates the edges in the subgraph induced on $\bS$. Only edges between two low-degree vertices ($\deg (u), \deg (v) \le k$) are counted.

\begin{algo}
    \procname{$\procnameZsampleZllZedgesZsparse(G,\bar m, k)$}
    \label{alg:e-ll-sparse}
    \alginput{Integer $n \geq 1$, parameters $k \ge 0$, $\mbar > 0$}
    \alginput{Access to the oracles \oracleIS, \oracleDeg of a graph $G=([n], E)$ with $m=|E|$}
    \algoutput{An unbiased estimation $\bX$, for $\frac{m_{LL}}{\bar m}$}
    \begin{code}
        \algitem Draw a subset $\bS \subseteq V$ that every vertex belongs to with probability $1 / \sqrt{\bar{m}}$, independently.
        \algitem Run $\procnameZenumerateZedges(G,\bS)$, and let $E_\bS$ be the resulting list of edges. \label{line:run-enumerate}
        \algitem Initialize $\bX \gets 0$.
        \begin{For}{each edge $(\bu,\bv) \in E_\bS$}
            \algitem Let $d_{\bu} \gets \deg (\bu)$ (via \oracleDeg).
            \algitem Let $d_{\bv} \gets \deg (\bv)$ (via \oracleDeg).
            \begin{If} {$d_{\bu} \le k$ and $d_{\bv} \le k$} \label{line:degudegv}
                \algitem Increment $\bX \gets \bX + 1$.
            \end{If}
        \end{For}
        \algitem Return $\bX$.
    \end{code}
\end{algo}

\begin{lemma} \label{lm:s-ll-sparse}
    The output of Procedure \procnameZsampleZllZedgesZsparseHREF (Algorithm \ref{alg:e-ll-sparse}) is a random variable whose expected value is $\frac{m_{LL}}{\bar{m}}$ and variance is at most $\left(1 + \frac{2k}{\sqrt{\bar{m}}}\right) \frac{m_{LL}}{\bar{m}}$. Moreover, the procedure's expected query complexity is $O(1 + \frac{m}{\bar m} \log n)$.
\end{lemma}
\begin{proof}
    For every edge $(u,v) \in E$, the marginal probability that both $u$ and $v$ belong to $\bS$ is $(1 / \sqrt{\bar{m}})^2 = 1 / \bar{m}$. Hence by linearity of expectation, the expected number of these edges is $m / \bar{m}$.
    
    Let $\bY$ be the number of edges in the subgraph induced on $\bS$'s vertices ($\bY = \abs{E_\bS}$). By Lemma \ref{lm:enumerate-edges}, the query complexity of \procnameZenumerateZedges~is $O(1 + \bY \log n)$. Every such an edge costs two additional degree queries, which sums to $2\bY$. Hence, the expected query complexity is $O(1 + \E[\bY] \log n) = O\left(1 + \frac{m}{\bar{m}} \log n\right)$.

    For every edge $e = (u,v)$ between two low-degree vertices, let $\bX_e$ be an indicator for $u,v \in \bS$. Observe that $\bX = \sum_{e \in E_{LL}} \bX_e$, and that $\Ex[\bX] = \frac{m_{LL}}{\bar{m}}$ since $\Ex[\bX_e] = 1 / \bar{m}$ for every edge $e$.

    Before we proceed to the variance, we bound the covariance of edge pairs. For edges $e_1 = (u_1,v_1)$ and $e_2 = (u_2,v_2)$ with no common vertex, the existence of $u_1,v_1$ in $\bS$ is independent of the existence of $u_2,v_2$, and therefore, $\Cov[\bX_{e_1}, \bX_{e_2}] = 0$. If $e_1 = (u,v_1)$ and $e_2 = (u,v_2)$ intersect in a single vertex, then $\Cov[\bX_{e_1}, \bX_{e_2}] = \Prx[u,v_1,v_2 \in \bS] - \Prx[u,v_1 \in \bS]\Prx[u,v_2 \in \bS] = \frac{1}{\bar{m}^{3/2}} - \frac{1}{\bar{m}^2}$.

    For the variance, recall that an edge in $E_{LL}$ can intersect at most $2k$ other edges in $E_{LL}$, since every vertex has degree at most $k$.
    \begin{eqnarray*}
        \Var[\bX]
        = \sum_{e \in E_{LL}} \left(\Var[\bX_e] + \sum_{\substack{e' \in E_{LL}\\e \cap e' \ne \emptyset}} \Cov[\bX_e,\bX_{e'}]\right)
        &\le& \sum_{e \in E_{LL}} \left(\frac{1}{\bar{m}} + 2k \cdot \frac{1}{\bar{m}^{3/2}}\right) \\
        &=& m_{LL} \left(\frac{1}{\bar{m}} + \frac{2k}{\bar{m}^{3/2}}\right)
        = \left(1 + \frac{2k}{\sqrt{\bar{m}}}\right) \frac{m_{LL}}{\bar{m}},
    \end{eqnarray*}
    completing the proof.
\end{proof}

In procedure \procnameZestimateZllZedgesZadviceHREF (Algorithm~\ref{alg:e-ll-advice}), we estimate $m_{LL}$ by averaging the outcomes of
$O\!\left(\frac{1 + k/\sqrt{\bar{m}}}{\eps^{2}}\right)$
invocations of \procnameZsampleZllZedgesZsparseHREF.
Since each invocation has expected value $m_{LL}/\bar{m}$, this averaging yields good concentration around the mean.

\begin{algo}
    \procname{$\procnameZestimateZllZedgesZadvice(n,\eps,G,\bar m, k)$}
    \label{alg:e-ll-advice}
    \alginput{Integer $n \geq 1$, parameters $\eps \in (0, 1], k \ge 0$, $\mbar > 0$}
    \alginput{Access to the oracles \oracleIS, \oracleDeg of a graph $G=([n], E)$ with $m=|E|$}
    \algoutput{A number $\tilde \boldm_{LL}$}
    \begin{code}
        \algitem Let $\bc\leftarrow 0$, $q \leftarrow \ceil{\frac{600}{\eps^2}\max\left(1,\frac{k}{\sqrt{\bar{m}}}\right)}$.
        \begin{For} {$i=1,...,q$}
            \algitem Run $\procnameZsampleZllZedgesZsparseHREF(n, G, \bar m, k)$ and let $\bX_i$ be its result.
            \algitem Increment $\bc \leftarrow \bc + \frac{1}{q}\bX_i$.
        \end{For}
        \algitem Return $\tilde \boldm_{LL} = \bar m\cdot \bc$.
    \end{code}
\end{algo}

\begin{lemma} \label{lm:e-ll-advice}
    Let $\tilde{\bm}_{LL}$ be the (random) output of procedure \procnameZestimateZllZedgesZadviceHREF (Algorithm \ref{alg:e-ll-advice}).
    \begin{itemize}
        \item If $\bar{m} \ge \frac{1}{4} m_{LL}$, then with probability at least $1 - \frac{m_{LL}}{200\bar{m}}$, $\abs{\tilde{\bm}_{LL} - m_{LL}} \le \eps \bar{m}$.
        \item If $\bar{m} < m_{LL}$, then with probability at least $1 - \frac{\bar{m}}{200m_{LL}}$, $\tilde{\bm}_{LL} > (1-\eps)\bar{m}$.
    \end{itemize}
    Moreover, the expected query complexity is $O\left(\frac{\log n}{\eps^2}\left(1 + \frac{k}{\sqrt{\bar m}}\right)\left(1 + \frac{m}{\bar{m}}\right)\right)$. Note that if $\frac{1}{4}m_{LL} \le \bar{m} < m_{LL}$, then both guarantees hold simultaneously.
\end{lemma}

\begin{proof}
    For query complexity, observe that by linearity of expectation and by Lemma \ref{lm:s-ll-sparse}, the expected query complexity is
    \[  q \cdot O\left(1 + \frac{m}{\mbar}\log n\right)
        = O\left(\frac{1}{\eps^2}\left(1 + \frac{k}{\sqrt{\mbar}}\right)\right) \cdot O\left(\left(1 + \frac{m}{\mbar}\right)\log n\right)
        = O\left(\frac{\log n}{\eps^2}\left(1 + \frac{k}{\sqrt{\mbar}}\right)\left(1 + \frac{m}{\mbar}\right)\right).
    \]

    We note that $\tilde \boldm_{LL}= \bar m \bc = \frac{\bar m}{q}\sum_{i=1}^{q} \bX_i$.
    The expected value of $\tilde \boldm_{LL}$, by linearity of expectation and Lemma \ref{lm:s-ll-sparse}:
    \[
    \Ex[\tilde \boldm_{LL}] = \frac{\bar m}{q}\sum_{i=1}^{q} \Ex[\bX_i] = \frac{\bar m}{q}\sum_{i=1}^{q} \frac{m_{LL}}{\bar m} = m_{LL}.
    \]
    
    The variance, as the $\bX_i$s are iid:
    \[
    \Varx[\tilde \boldm_{LL}] = \left(\frac{\bar m}{q}\right)^2\sum_{i=1}^{q}\Varx[\bX_i] = \frac{\bar m^2}{q}\Varx[\bX_1].
    \]
    
    By Lemma \ref{lm:s-ll-sparse} we can bound:
    \[
    \Varx[\tilde \boldm_{LL}] \leq \frac{\bar m^2}{q}\left(1+ \frac{2k}{\sqrt{\bar m}}\right)\frac{m_{LL}}{\bar m} = \frac{m_{LL}\cdot\bar m}{q}\left(1+ \frac{2k}{\sqrt{\bar m}}\right)
    \]
    
    By Chebyshev's inequality:
    \begin{eqnarray*}
        \Prx[|\tilde \boldm_{LL} - m_{LL}| \geq \eps \bar m] \leq \frac{\Varx[\tilde \boldm_{LL}]}{\eps^2 \bar m^2} &\leq& \frac{\frac{m_{LL}\cdot\bar m}{q}\left(1+ \frac{2k}{\sqrt{\bar m}}\right)}{\eps^2 \bar m^2} = \frac{m_{LL}\left(1+ \frac{2k}{\sqrt{\bar m}}\right)}{\eps^2\cdot q\cdot\bar m}\\
        &=& \frac{m_{LL}}{\eps^2\cdot q\cdot\bar m} + \frac{m_{LL}\frac{2k}{\sqrt{\bar m}}}{\eps^2\cdot q\cdot\bar m}\\
        &\leq& \frac{m_{LL}}{\eps^2\cdot \frac{600}{\eps^2}\cdot \bar m} + \frac{m_{LL}\frac{2k}{\sqrt{\bar m}}}{\eps^2\cdot \frac{600}{\eps^2}\cdot \frac{k}{\sqrt{\bar m}}\cdot\bar m}\\
        &=& \frac{m_{LL}}{600\bar m} + \frac{2m_{LL}}{600\bar m} = \frac{m_{LL}}{200\bar m}.
    \end{eqnarray*}
    
    Thus, when $\bar m \ge \frac{1}{4} m_{LL}$, with probability at least $1-\frac{m_{LL}}{200\bar m}$, $\tilde \boldm_{LL}$ satisfies $m_{LL} - \eps \bar m \leq \tilde \boldm_{LL} \leq m_{LL} + \eps \bar m$, satisfying the first condition of the lemma.

    If $\bar m < m_{LL}$, we use the following bound:
    \[
        \Prx[\tilde \boldm_{LL} \le (1-\eps)\bar m] \leq \Prx[\tilde \boldm_{LL} \leq (1-\eps) m_{LL}]
        = \Prx\left[\tilde \boldm_{LL} \leq \Ex[\tilde \boldm_{LL}] - \eps m_{LL}.\right]
    \]

    By Chebyshev's inequality:
    \begin{eqnarray*}
        \Prx[\tilde \boldm_{LL} \le (1-\eps)\bar m] &\leq& \frac{\Varx[\tilde \boldm_{LL}]}{\eps^2 m_{LL}^2} \leq \frac{\frac{m_{LL}\cdot\bar m}{q}\left(1+ \frac{2k}{\sqrt{\bar m}}\right)}{\eps^2m_{LL}^2}\\
        &=& \frac{\bar m}{m_{LL}}\cdot \frac{\left(1+ \frac{2k}{\sqrt{\bar m}}\right)}{\eps^2q}\\
        &=& \frac{\bar m}{m_{LL}}\cdot \left(\frac{1}{\eps^2q} +\frac{\frac{2k}{\sqrt{\bar m}}}{\eps^2q} \right)\\
        &\leq& \frac{\bar m}{m_{LL}}\cdot \left(\frac{1}{600} +\frac{2}{600} \right)\\
        &=& \frac{\bar m}{200m_{LL}}.
    \end{eqnarray*}

    Hence, when $\bar m < m_{LL}$, with probability at least $1-\frac{\bar m}{200m_{LL}}$, $\tilde \boldm_{LL}$ satisfies $(1-\eps)\bar m\leq \tilde \boldm_{LL}$, satisfying the second condition of the lemma.
\end{proof}

\subsection{Estimating $m_{LH}$ (with advice)}

In this subsection we provide an estimation algorithm for the number of edges between low-degree vertices ($\deg (u) \le k$) and high-degree vertices ($\deg (u) > k$). Recall that the number of low-high edges is denoted by $m_{LH}$, and the set of these edges is denoted by $E_{LH}$.

We define the ``very-low'' degree threshold as $k' = \min (k, \bar{m} / \eps k )$. Let $V_{L_1}$ denote the set of vertices of degree at most $k'$, let $E_{L_1H}$ be the set of edges between vertices in $V_{L_1}$ and high-degree vertices, and let $m_{L_1H} = |E_{L_1H}|$.
Since $V_{L_1} \subseteq V_L$, it follows that $m_{L_1H} \le m_{LH}$.
Moreover, Lemma \ref{lm:just-few-edges-above-kprime-to-high} implies that the number of edges between high-degree vertices and vertices of degree exceeding $k'$ is at most $4\eps\bar m$. Consequently, $m_{L_1H} \le m_{LH} \le m_{L_1H} + 4 \eps \bar{m}$.

The estimator for $m_{L_1H}$, implemented as procedure \procnameZestimateZLoneHZedgesZadviceHREF (Algorithm \ref{alg:e-l1h-advice}), averages over $O(nk' / \eps^2 \bar{m})$ samples as follows: we uniformly draw a vertex $\bu$ and a neighbor $\bv$ of $\bu$ (if $\deg (\bu) = 0$, we proceed to the next sample). If $\bu$'s degree is ``very-low" (i.e., $\deg (\bu) \le k'$) and $\bv$'s degree is high ($\deg( \bv) > k$), then we add $\deg (\bu)$ to a counter $\bc$. For all other cases, we do not increment the counter. The result is normalized by an $n$ factor.

\begin{algo}
    \procname{$\procnameZestimateZLoneHZedgesZadvice(n,\eps,G,\bar m, k)$}
    \label{alg:e-l1h-advice}
    \alginput{Integers $n \geq 1$, parameters $\eps \in (0, 1], k \ge 0$, $\mbar > 0$}
    \alginput{Access to the oracles \oracleDeg, \oracleN of a graph $G=([n], E)$ with $m=|E|$}
    \algoutput{An unbiased estimator $\tilde \boldm_{L_1H}$ for $m_{L_1H}$}
    \begin{code}
        \algitem Initialize $\bc \gets 0$.
        \algitem Let $k' \gets \min\left( k, \frac{\bar m}{\eps k}\right)$.
        \algitem Let $t \gets \ceil{\frac{200nk'}{\eps^2\bar m}}$.
        \begin{For} {$i=1,...,t$}
            \algitem Sample a vertex $\bu \in V$ uniformly at random.
            \algitem Sample a neighbor $\bv$ of $\bu$ uniformly at random (via \oracleN).
            \algitem Let $\bd_u \gets \deg (\bu)$ (via \oracleDeg).
            \algitem Let $\bd_v \gets \deg(\bv)$ (via \oracleDeg).
            \begin{If} {$\bd_u \le k'$ and $\bd_v > k$}
                \algitem Increment $\bc \gets \bc + \bd_u$. \label{line:e-l1h-increment}
            \end{If}
        \end{For}
        \algitem Return $\tilde \boldm_{L_1H} = \frac{n}{t}\bc$.
    \end{code}
\end{algo}

\begin{lemma} \label{lm:e-l1h-advice}
    Let $\tilde{\bm}_{L_1,H}$ be the (random) output of \procnameZestimateZLoneHZedgesZadviceHREF (Algorithm \ref{alg:e-l1h-advice}) when executed with parameters $(G,\eps,\bar{m},k)$. 
    \begin{itemize}
        \item If $\bar{m} \ge \frac{1}{4}m_{L_1H}$, then with probability at least $1 - \frac{m_{L_1H}}{200\bar{m}}$, $\abs{\tilde{\bm}_{L_1H} - m_{L_1H}} \le \eps \bar{m}$.
        \item If $\bar{m} < m_{L_1H}$, then with probability at least $1 - \frac{\bar{m}}{200m_{L_1H}}$, $\tilde{\bm}_{L_1H} > (1 - \eps) \bar{m}$.
    \end{itemize}
    Moreover, the query complexity is bounded by $O(n / \eps^3 k)$.
\end{lemma}

\begin{proof}
    For the query complexity, observe that every iteration makes exactly four queries, and that the number of iterations is $O(1) \cdot \frac{nk'}{\eps^2 \bar{m}} \le O(1) \cdot \frac{n (\bar{m} / \eps k)}{\eps^2 \bar{m}} = O(n / \eps^3 k)$.

    For every $1 \le i \le t$, let $\bX_i$ be the value added to $\bc$ (line \ref{line:e-l1h-increment}), or zero if nothing was added. The output of the algorithm is $n \bar{\bX}$, where $\bar{\bX} = \frac{1}{t}\sum_{i=1}^t \bX_i$.

     For a vertex $v$ we denote by $\deg_H(v)$ the number of neighbors of $v$ which are high-degree. Thus, the expected value
    \begin{eqnarray*}
        \Ex[\bar{\bX}] = \Ex[\bX_1] &=& 
        \sum_{u\; :\; \deg (u) \le k'} \Prx[\bu = u] \cdot \Prx_{\bv}\left[\deg (\bv) > k \mid \bu = u \right] \cdot \deg (u) \\
        &=& \sum_{u \in V_{L_1}} \frac{1}{n} \cdot \frac{\deg_H (u)}{\deg (u)} \cdot \deg (u)= \frac{1}{n} \sum_{u \in V_{L_1}} \deg_H (u)
        = \frac{m_{L_1H}}{n}.
    \end{eqnarray*}
    
    For the second moment:
    \begin{eqnarray*}
        \Ex[\bX_1^2] &=& \sum_{u \in V_{L_1}} \Prx[\bu = u] \cdot \Prx\left[\deg (\bv) > k \mid \bu = u \right] \cdot (\deg (u))^2 \\
        &=& \sum_{u \in V_{L_1}} \frac{1}{n} \cdot \frac{\deg_H (u)}{\deg (u)} \cdot \deg (u)^2 = \frac{1}{n} \sum_{u \in V_{L_1}} \deg_H (u) \cdot \deg (u) \\
        &\le& \frac{k'}{n} \sum_{u \in V_{L_1}} \deg_H (u)
        = k' \cdot\Ex[\bX_1].
    \end{eqnarray*}
    
    For the variance:
    \[  \Var[\bX_1]
        \le \Ex[\bX_1^2]
        \le k' \Ex[\bX_1]
        = k' \frac{m_{L_1,H}}{n}.\]

    Since the $\bX_i$'s are independent and identically distributed,
    \[  \Var[\bar{\bX}]
        \le \frac{k'}{t}\E[\bar{\bX}]
        \le \frac{k'}{200 nk' / \eps^2 \bar{m}} \cdot \frac{m_{L_1 H}}{n}
        = \frac{\eps^2 \bar{m} m_{L_1 H}}{200 n^2}
    \]

    By Chebyshev's inequality, if $\bar{m} \ge \frac{1}{4}m_{L_1 H}$, then:
    \[  \Prx\left[\abs{\tilde \boldm_{L_1H}-m_{L_1H}} \ge \eps \bar{m} \right]
        = \Prx\left[\abs{\bar{\bX} - \E[\bar{\bX}]} \ge \frac{\eps}{n}\bar{m} \right]
        \le \frac{\Var[\bar{\bX}]}{\eps^2 \bar m^2 / n^2}
        \le n^2 \frac{\eps^2 m_{L_1 H} \bar{m} / 200n^2}{\eps^2 \bar {m}^2}
        = \frac{m_{L_1 H}}{200 \bar {m}}. \]

    If $\bar{m} < m_{L_1H}$, then:
    \[  \Prx\left[\tilde \boldm_{L_1H} \!\le\! (1 - \eps) \bar{m} \right]
        \le \Prx\left[\bar{\bX} \!\le\! (1 - \eps)\E[\bar{\bX}] \right]
        \le \frac{\Var[\bar{\bX}]}{\eps^2 m_{L_1H}^2 / n^2}
        \le n^2 \frac{\eps^2 m_{L_1H} \bar{m} / 200n^2}{\eps^2 m_{L_1H}^2}
        = \frac{\bar{m}}{200 m_{L_1H}}. \]
\end{proof}

\subsection{Handling the case where $\mbar < m$}

Beame's algorithm (\procnameZenumerateZedges) is meant to be executed as a whole. For our algorithm, we need a more fine-grained analysis. We prove the following lemma in Appendix \ref{apx:enumerate-edges-amortized}:
\begin{lemma} \label{lm:enumerate-edges-amortized}
    \procnameZenumerateZedges (Lemma \ref{lm:enumerate-edges}) can be implemented with an additional guarantee: for every $1 \le t \le \abs{E_A}$, the $t$-th edge is added to the list after performing at most $O(1 + t \log n)$ queries. In other words, it has $O(1)$ initialization phase and amortized $O(\log n)$ ``next-element'' phase.
\end{lemma}

Procedure \procnameZsmallZmbarZguardHREF (Algorithm~\ref{fig:alg:small-mbar-guard}) separates the regime in which the advice $\bar{m}$ is valid ($\bar{m} \ge m$) from the regime in which it is substantially underestimated ($\mbar < m/4$) or when it is structurally wrong ($\max_u \deg (u) > \mbar$). In the former case, the procedure accepts with probability at least $1 - 1/1000$ (completeness), while in the latter it rejects with probability at least $1 - \bar{m}/1000m$ (soundness). The procedure searches for vertices whose degree exceeds $\bar{m}$ and applies a moderated variant of \procnameZsampleZllZedgesZsparseHREF in order to reject instances in which the number of edges is significantly larger than what is consistent with the assumption $\bar{m} \ge m$.

For readability, we provide a parameterized algorithm and a corresponding parameterized analysis for completeness $1 - c$ and soundness $1 - c\mbar / m$, where $c \le 1$. For the lemma statement, we use $c=1/1000$.

\paragraph{Case 0 (trivial)} $m = O(\poly(1/c))$. This is an easy case since we can just look for $O(\poly(1/c))$ edges at the cost of $O_c(\log n)$ queries to find out whether or not $\mbar \ge \min(m, \poly(1/c)) = m$.

\paragraph{Case I (typical)} $m_{LL} \ge \frac{3}{4}m$, where the threshold degree $k$ is defined as $\min(\mbar, n-1)$.

\paragraph{Case II (very-high degree)} There exists a vertex of degree at least $\sqrt{\mbar} \ln (m/c)$.

\paragraph{Case III (sparse)} $m = \Omega(\poly(1/c))$ and $m_{LL} < \frac{3}{4}m$ (where the threshold degree $k$ is defined as $\min(\mbar, n-1)$) and the maximum degree is at most $\sqrt{\mbar} \ln (m/c)$.

Observe that Case I is ``quantitative'' (to be resolved by large-deviation inequalities), whereas Case II is a ``qualitative'' (to be resolved by looking for a witness), and therefore, they are treated using different algorithms. Case III is analyzed through the qualitative algorithm resolving Case II.

\subsubsection{Quantity test (Case I)}
In the quantity test (\procnameZguardZquantityHREF, Algorithm \ref{fig:alg:small-mbar-guard-APX-CASE-QUANTITY}) we execute an edge sampler similar to \procnameZestimateZllZedgesZadviceHREF (Algorithm \ref{alg:e-ll-advice}), but we use the lazy variant of \procnameZenumerateZedges (Lemma \ref{lm:enumerate-edges-amortized}) to keep the query complexity low even if $\mbar$ is significantly smaller than $m$. Additionally, if we happen to encounter a vertex whose degree is greater than $\mbar$, then we reject immediately.

\begin{algo}
    \procname{$\procnameZguardZquantity(c;G,\bar{m})$}
    \label{fig:alg:small-mbar-guard-APX-CASE-QUANTITY}
    \algcomplexity{$O(\frac{1}{c} \sqrt{\bar{m}} \log n)$}
    \algcompleteness{If $\bar{m} \ge m$, then we accept with probability at least $1 - c/2$}
    \algsoundness{If $\bar{m} < \frac{1}{4}m$ and $m_{LL} \ge \frac{3}{4}m$, then we reject with probability at least $1 - \frac{c\bar{m}}{m}$}
    \algsoundness{If $m_{LL} < \frac{1}{2}m$ and the maximum degree in $G$ is less than $\sqrt{\mbar} \ln (m/c)$, then we reject with probability at least $1 - \frac{c\bar{m}}{m}$}
    \begin{code}
        \algitem Initialize $\bX \gets 0$.
        \algitem Let $t \gets \ceil{\frac{96}{c} \sqrt{\bar{m}}}$.
        \begin{For}{$t$ times}
            \algitem Draw a set $\bS \subseteq V$ that every vertex belongs to with probability $\frac{1}{\sqrt{\bar{m}}}$, iid.
            \algitem Initialize \procnameZenumerateZedges, let $L_\bS$ be the result, accessible sequentially.
            \begin{While}{$L_{\bS}$ has an unfeteched edge}
                \algitem Fetch $(\bu,\bv) \in L_{\bS}$.
                \algitem Let $d_{\bu} \gets \deg (\bu)$ via \oracleDeg query.
                \algitem Let $d_{\bv} \gets \deg (\bv)$ via \oracleDeg query.
                \begin{If}{$d_{\bu} > \bar{m}$ or $d_{\bv} > \bar{m}$}
                    \algitem Return \textsc{reject}.\label{fig:alg:small-mbar-guard-APX-CASE-II-III::line:rej-by-high-degree}
                \end{If}
                \algitem Increment $\bX \gets \bX + 1$.
                \begin{If}{$\bX \ge (5/4)t$}
                    \algitem Return \textsc{reject}.
                \end{If}
            \end{While}
        \end{For}
        \algitem Return \textsc{accept}.
    \end{code}
\end{algo}

\begin{observation}
    Considering Algorithm \ref{fig:alg:small-mbar-guard-APX-CASE-QUANTITY} (\procnameZguardZquantityHREF), ignoring the possibility of an early termination (but without counting the high-degree-adjacent edges that would cause it), the gain in $\bX$ in every iteration distributes the same as in Algorithm \ref{alg:e-ll-sparse} (\procnameZsampleZllZedgesZsparseHREF).
\end{observation}

\begin{observation} \label{obs:guard-quantity-iteration-gain-X}
    Assume that $\mbar \ge 1$. Considering Algorithm \ref{fig:alg:small-mbar-guard-APX-CASE-QUANTITY} (\procnameZguardZquantityHREF), ignoring the possibility of an early termination (but without counting the high-degree-adjacent edges that would cause it), the gain in $\bX$ in every iteration has expected value $\frac{m_{LL}}{\mbar}$ and its variance is bounded by $3\sqrt{\mbar} \cdot \frac{m_{LL}}{\mbar}$.
\end{observation}

\begin{lemma}[Completeness of $\procnameZguardZquantityHREF$] \label{lm:guard-quantity-completeness}
    If $\mbar \ge m$, then $\procnameZguardZquantityHREF$ accepts with probability at least $1 - \frac{1}{2}c$.
\end{lemma}
\begin{proof}
    Since the graph is assumed to have edges, $m \ge 1$, and therefore, $\mbar \ge 1$.
    
    The probability to reject due to finding a vertex of degree greater than $\mbar$ is zero, since the maximum degree in the graph is bounded by $m \le \mbar$.

    By Observation \ref{obs:guard-quantity-iteration-gain-X}, ignoring the possibility of early termination, $\Ex[\bX] = \frac{m_{LL}}{\mbar} t \le t$ and $\Varx[\bX] \le 3\sqrt{\mbar} t$. Therefore, by Chebyshev's inequality, the probability to reject is:
    \begin{eqnarray*}
        \Pr\left[\bX \ge (5/4)t\right]
        \le \Pr\left[\bX \ge \E[\bX] + t/4\right]
        \le \frac{3\sqrt{\mbar}t}{(1/16)t^2}
        =  48 \cdot \frac{\sqrt{\mbar}}{t}
        \le \frac{48}{96/c}
        = \frac{1}{2}c.
    \end{eqnarray*}
\end{proof}

\begin{lemma}[Soundness of $\procnameZguardZquantityHREF$ wrt Case I] \label{lm:guard-quantity-soundness-I}
    If $1 \le \mbar < \frac{1}{4}m$ and $m_{LL} \ge \frac{3}{4}m$, where the threshold degree $k$ is defined as $\min(\mbar, n-1)$, then \procnameZguardZquantityHREF rejects with probability at least $1 - \frac{c\bar{m}}{m}$. 
\end{lemma}
\begin{proof}
    By Observation \ref{obs:guard-quantity-iteration-gain-X}, ignoring the possibility of early termination, $\Ex[\bX] = \frac{m_{LL}}{\mbar} t \ge \frac{(3/4)m}{\mbar}t \ge 3t$ and $\Varx[\bX] \le 3\sqrt{\mbar} \E[\bX]$. Therefore, by Chebyshev's inequality, the probability to accept is bounded by:
    \begin{eqnarray*}
        \Pr\left[\bX < (5/4)t\right]
        \le \Pr\left[\bX < \frac{1}{2}\Ex[\bX]\right]
        &\le& \frac{3\sqrt{\mbar}\Ex[\bX]}{(1/4)(\E[\bX])^2} \\
        &=& 12 \cdot \frac{\sqrt{\mbar}}{\E[\bX]}
        \le 12 \cdot \frac{\sqrt{\mbar}}{((3/4)m / \mbar) \cdot t}
        \le \frac{16}{96/c} \cdot \frac{\mbar}{m}
        \le \frac{c \mbar}{m}.
    \end{eqnarray*}
    (Additionally, the procedure can also reject by encountering an edge adjacent to a high-degree vertex)
\end{proof}

\begin{lemma} \label{lm:guard-quantity-complexity}
    The query complexity of $\procnameZguardZquantityHREF$ is $O(\frac{1}{c} \cdot \sqrt{\mbar} \log n)$ worst-case.
\end{lemma}
\begin{proof}
    By Lemma \ref{lm:enumerate-edges-amortized}, the query complexity of \procnameZguardZqualityHREF is $O(t + \bX \log n)$. Since the procedure aborts once $\bX$ reaches $(5/4)t$, the worst-case query complexity is bounded by $O(t \log n) = O(\frac{1}{c} \sqrt{\mbar} \log n)$.
\end{proof}

\subsubsection{Quality test (Case II, Case III)}
In the quality test (\procnameZguardZqualityHREF, Algorithm \ref{fig:alg:small-mbar-guard-APX-CASE-QUALITY}) we use a similar logic to the quantitative test, but instead of having independent iterations, we choose the sets such that every vertex belongs to exactly one iteration. This guarantees that every potential witness belongs to the union of the iteration sets, which is not guaranteed in Algorithm \ref{fig:alg:small-mbar-guard-APX-CASE-QUANTITY}. We reject as soon as we encounter a vertex whose degree is strictly greater than $\mbar$, or when we find significantly more edges than what we expect to if we would run \procnameZestimateZllZedgesZadviceHREF.

\begin{algo}
    \procname{$\procnameZguardZquality(c;G,\bar{m})$}
    \label{fig:alg:small-mbar-guard-APX-CASE-QUALITY}
    \algcomplexity{$O(\sqrt{\bar{m}} \log n)$}
    \algcompleteness{If $\bar{m} \ge m$, then we accept with probability at least $1 - c/2$}
    \algsoundness{If $\max_u \deg (u) \ge \sqrt{\mbar} \ln (m/c)$, then we reject with probability at least $1 - \frac{c\mbar}{m}$}
    \begin{code}
        \algitem Let $t \gets \ceil{\sqrt{\mbar}}$.
        \begin{For}{each $u \in V$}
            \algitem Draw $\bi_u \in \{1,\ldots,t\}$ uniformly and independently.
        \end{For}
        \algitem Initialize $\bX \gets 0$.
        \begin{For}{$i$ from $1$ to $t$}
            \algitem Let $\bS_i \gets \{ u \in V : \bi_u = i \}$.
            \algitem Initialize \procnameZenumerateZedges, let $L_\bS$ be the result, accessible sequentially.
            \begin{While}{$L_{\bS}$ has an unfeteched edge}
                \algitem Fetch $(\bu,\bv) \in L_{\bS}$.
                \algitem Let $d_{\bu} \gets \deg (\bu)$ via \oracleDeg query.
                \algitem Let $d_{\bv} \gets \deg (\bv)$ via \oracleDeg query.
                \begin{If}{$d_{\bu} > \bar{m}$ or $d_{\bv} > \bar{m}$}
                    \algitem Return \textsc{reject}.
                \end{If}
                \algitem Increment $\bX \gets \bX + 1$.
                \begin{If}{$\bX \ge 2t/c$}
                    \algitem Return \textsc{reject}.
                \end{If}
            \end{While}
        \end{For}
        \algitem Return \textsc{accept}.
    \end{code}
\end{algo}

\begin{observation} \label{obs:guard-quality-iteration-gain-X}
    Assume that $\mbar \ge 1$. Considering Algorithm \ref{fig:alg:small-mbar-guard-APX-CASE-QUANTITY} (\procnameZguardZquantityHREF), ignoring the possibility of an early termination (but without counting the high-degree-adjacent edges that would cause it), the gain in $\bX$ in every iteration has expected value $\frac{m_{LL}}{\mbar}$ and its variance is bounded by $3\sqrt{\mbar} \cdot \frac{m_{LL}}{\mbar}$.
\end{observation}

\begin{lemma}[Completeness of $\procnameZguardZqualityHREF$] \label{lm:guard-quality-completeness}
    If $\mbar \ge m$, then $\procnameZguardZqualityHREF$ accepts with probability at least $1 - \frac{1}{2}c$.
\end{lemma}
\begin{proof}
    The probability to reject due to finding a vertex of degree greater than $\mbar$ is zero, since the maximum degree in the graph is bounded by $m \le \mbar$.

    By Observation \ref{obs:guard-quality-iteration-gain-X}, the expected value of $\bX$ is bounded by $\frac{m}{\mbar} t \le t$. Therefore, by Markov's inequality, $\Pr\left[\bX \ge 2t/c\right] \le \frac{1}{2}c$.
\end{proof}

\begin{lemma}[Soundness of \procnameZguardZqualityHREF wrt Case II] \label{lm:guard-quality-soundness-II}
    Assume that $\mbar \ge 1$. If there exists a vertex whose degree is at least $\sqrt{\mbar} \ln (m/c)$, then \procnameZguardZqualityHREF rejects with probability at least $1 - \frac{c\mbar}{m}$.
\end{lemma}
\begin{proof}
    Let $u^*$ be an arbitrary vertex of degree at least $D = \sqrt{\mbar} \ln (m/c)$, and consider $\bS = S_{\bi_{u^*}}$. The probability to not-reject in the $\bi_{u^*}$-th iteration is bounded by the probability that $i_v \ne i_{u^*}$ for every neighbor $v$ of $u^*$. Therefore, the probability to accept in all iterations is bounded by:
    \[  \left(1 - \frac{1}{\sqrt{\mbar}}\right)^{\deg (u^*)}
        \le \left(1 - \frac{1}{\sqrt{\mbar}}\right)^{\sqrt{\mbar} \ln (m/c)}
        \le e^{-\ln (m/c)}
        = \frac{c}{m}
        \le \frac{c\mbar}{m}.
    \]
\end{proof}

\begin{lemma}[Soundness of $\procnameZguardZquantityHREF$ wrt Case III] \label{lm:guard-quality-soundness-III}
    Assume that $m \ge \max(16(1 + \ln c^{-1})^{14}, 1/c)$. If $1 \le \mbar < m$ and $m_{LL} < \frac{3}{4}m$, where the threshold degree $k$ is defined as $\min(\mbar, n-1)$, and the maximum degree in the graph is bounded by $\sqrt{\mbar} \ln (m/c)$, then the probability that \procnameZguardZqualityHREF accepts is bounded by least $\frac{c\bar{m}}{m}$.
\end{lemma}
\begin{proof}
    Let $D = \sqrt{\mbar} \ln (m/c)$ be an upper bound for the maximum degree in the graph.

    Since $m_{LL} < m$, we deduce that $k$ is smaller than the maximum degree in the graph, and therefore, $k = \mbar$ (since it cannot be equal to $n-1$). Let $H = \{ u \in V : \deg (u) > \mbar \}$ be the set of vertices whose degree is greater than $k = \min(\mbar, n-1)$. Note that this set is not empty since $m_{LL} < m$. Since $m_{LL} < \frac{3}{4}m$, $\abs{H} \ge \frac{m - m_{LL}}{2D} > \frac{m/4}{2D} = \frac{m}{8D}$.

    We define two bounds: $\mbar_\mathrm{lbnd} = \ln^2 (m/c)$ and $\mbar_\mathrm{ubnd} = \left(\frac{1}{16} m / \ln^4 (m/c)\right)^{2/3}$.

    If $\mbar \ge \mbar_\mathrm{lbnd} = \ln^2 (m/c)$, then for an arbitrary $u^* \in H$, the probability that $\bS_{\bi_{u^*}}$ has no $u^*$-neighbor is bounded by:
    \begin{eqnarray*}
        \left(1 - \frac{1}{\sqrt{\mbar}}\right)^{\mbar}
        \le e^{-\sqrt{\mbar}}
        \le e^{-\ln (m/c)}
        \le \frac{c}{m}
        \le \frac{c\mbar}{m}.
    \end{eqnarray*}

    Now we consider lower values of $\mbar$. Let $U \subseteq H$ be an independent set of high-degree vertices that have no common neighbor. If we construct $U$ greedily, then $\abs{U} \ge \frac{\abs{H}}{1 + D + D(D-1)} \ge \frac{\abs{H}}{2D^2} \ge \frac{m}{16D^3}$, since every vertex has at most $D$ neighbors and $D(D-1)$ neighbors-of-neighbors.

    Since $U$ is independent and its vertices have pairwise-disjoint neighbor sets, the events of the form ``there exists an edge adjacent to $u$'' (for $u \in U$) are independent.

    If $\mbar \le \mbar_\mathrm{ubnd} = \left(\frac{1}{16} m / \ln^4 (m/c)\right)^{2/3}$, then
    \begin{eqnarray*}
        \abs{U}
        \ge \frac{m}{16D^3}
        &\ge& \frac{m}{16 (\sqrt{\mbar} \ln (m/c))^3} \\
        &\ge& \frac{m}{16 \left(\sqrt{\left((1/16) m / \ln^4 (m/c)\right)^{2/3}}\right)^3 \ln^3 (m/c)} \\
        &=& \frac{m}{16 \left((1/16) m / \ln^4 (m/c)\right) \ln^3 (m/c)}
        = \ln (m/c).
    \end{eqnarray*}
    
    The probability that for every $u \in U$, all its neighbors are outside $\bS_{\bi_u}$, is bounded by:
    \[  \prod_{u\in U} \left(1 - \frac{1}{t}\right)^{\deg (u)}
        < \prod_{u\in U} \left(1 - \frac{1}{\sqrt{\mbar}}\right)^{\mbar}
        \le \prod_{u\in U} \left(1 - \frac{1}{\sqrt{\mbar}}\right)^{\sqrt{\mbar}}
        \le e^{-\abs{U}}
        \le e^{-\ln (m/c)}
        = \frac{c}{m}
        \le \frac{c\mbar}{m}.
    \]

    Observe that $\mbar_{\mathrm{lbnd}} \le \mbar_{\mathrm{ubnd}}$ since $m \ge \max(16(1 + \ln c^{-1})^{14}, 1/c)$, and therefore, the analysis holds for every $1 \le \mbar < m$ for which $m_{LL} < \frac{3}{4}m$ and $\max_u \deg (u) \le \sqrt{\mbar} \ln (m/c)$.
\end{proof}

\begin{lemma}[Complexity of \procnameZguardZqualityHREF] \label{lm:guard-quality-complexity}
    If $\mbar \ge 1$, then the query complexity of \procnameZguardZqualityHREF is $O(\frac{1}{c} \sqrt{\mbar} \log n)$ worst-case.
\end{lemma}
\begin{proof}
    By Lemma \ref{lm:enumerate-edges-amortized}, the query complexity of \procnameZguardZqualityHREF is $O(t + \bX \log n)$. Since the procedure aborts once $\bX$ reaches $2t/c = O(\sqrt{m}/c)$, the worst-case query complexity is bounded by $O(\frac{1}{c} \sqrt{\mbar} \log n)$.
\end{proof}

\subsubsection{The guard procedure}

The guard procedure is merely a conjunction of a deterministic fixed-count edge enumeration (to detect Case 0), the quantitative guard procedure (to detect Case I) and the qualitative procedure (to detect Case II and Case III).

\begin{algo}
    \procname{$\procnameZsmallZmbarZguard(G,\bar{m})$}
    \label{fig:alg:small-mbar-guard}
    \algcomplexity{$O(\sqrt{\bar{m}} \log n)$}
    \algcompleteness{If $\bar{m} \ge m$ and $\max_u \deg (u) \le \bar{m}$, then we accept with probability at least $1 - \frac{1}{1000}$}
    \algsoundness{If $\bar{m} < \frac{1}{4}m$, then we reject with probability at least $1 - \frac{\bar{m}}{1000m}$}
    \algsoundness{If $\max_u \deg (u) > \mbar$, then we reject with probability at least $1 - \frac{\bar{m}}{1000m}$}
    \begin{code}
        \algitem Let $c \gets 1/1000$.
        \algitem Let $\mbar^* \gets \max(16(1 + \ln c^{-1})^{14}, 1/c)$.
        \algitem Initialize \procnameZenumerateZedges for the whole graph, let $L$ be the result list, accessible sequentially.
        \algitem Initialize $X \gets 0$.
        \begin{While}{$X \le \mbar^*$ and $L$ has an unfeteched edge}
            \algitem Consume an arbitrary edge of $L$.
            \algitem Increment $X \gets X + 1$.
        \end{While}
        \begin{If}{$X > \mbar$}
            \algitem Return \textsc{reject}.
        \end{If}
        \algitem Run $\procnameZguardZquantityHREF(c;G,\mbar)$.
        \begin{If}{The quantity guard rejects}
            \algitem Return \textsc{reject}.
        \end{If}
        \algitem Run $\procnameZguardZqualityHREF(c;G,\mbar)$.
        \begin{If}{The quality guard rejects}
            \algitem Return \textsc{reject}.
        \end{If}
        \algitem Return \textsc{accept}.
    \end{code}
\end{algo}

\begin{lemma} \label{lm:guard-new}
    Procedure \procnameZsmallZmbarZguard (Algorithm \ref{fig:alg:small-mbar-guard}) has:
    \begin{itemize}
        \item Completeness: if $\mbar \ge m$, then it accepts with probability at least $1 - \frac{1}{1000}$.
        \item Soundness: if $\mbar < \frac{1}{4}m$ or if the maximum degree in the graph is greater than $\mbar$, then it rejects with probability at least $1 - \frac{\mbar}{1000m}$.
        \item Complexity: the query complexity is $O(\sqrt{\mbar} \log n)$.
    \end{itemize}
\end{lemma}
\begin{proof}
    In all following lemmas we use $c = 1/1000$ (as hard-coded into Algorithm \ref{fig:alg:small-mbar-guard}).
    
    For completeness, observe that we cannot reject by $X > \mbar$, since $X \le m \le \mbar$ with probability $1$. Each of the guard subroutines has completeness $1 - \frac{1}{2}c$ (Lemma \ref{lm:guard-quantity-completeness}, Lemma \ref{lm:guard-quality-completeness}), and hence, by the union bound, the completeness is $1 - c = 1 - \frac{1}{1000}$.

    For soundness, consider the following cases:
    \begin{itemize}
        \item If $m \le \mbar^* = \max(16(1+\ln c^{-1})^{14}, 1/c)$ and $\mbar < m$ (Case 0), then the deterministic part of the algorithm causes reject (Lemma \ref{lm:enumerate-edges-amortized}).
        \item If $m_{LL} \ge \frac{3}{4}m$ (Case I), then \procnameZguardZquantityHREF rejects with probability at least $1 - \frac{c\mbar}{m} = 1 - \frac{\mbar}{1000m}$ (Lemma \ref{lm:guard-quantity-soundness-I}).
        \item If $m_{LL} < \frac{3}{4}m$ and there exists a vertex of degree at least $\sqrt{\mbar} \ln (m/c)$ (Case II), then \procnameZguardZqualityHREF rejects with probability at least $1 - \frac{c\mbar}{m} = 1 - \frac{\mbar}{1000m}$ (Lemma \ref{lm:guard-quality-soundness-II}).
        \item If $m > \mbar^*$ and $m_{LL} < \frac{3}{4}m$ and the maximum degree in the graph is greater than $\mbar$ and smaller than $\sqrt{\mbar} \ln (m/c)$ (Case III), then \procnameZguardZqualityHREF rejects with probability at least $1 - \frac{c\mbar}{m} = 1 - \frac{\mbar}{1000m}$ (Lemma \ref{lm:guard-quality-soundness-III}).
    \end{itemize}

    For complexity, observe that both subroutines cost $O(\sqrt{\mbar} \log n)$ (Lemma \ref{lm:guard-quantity-complexity}, \ref{lm:guard-quality-complexity}). The deterministic part costs $O(\mbar^* \log n)$ queries (Lemma \ref{lm:enumerate-edges-amortized}), which is $O(\log n)$ since $\mbar^*$ is a constant. Combined, the query complexity is bounded by $O(\sqrt{\mbar} \log n)$.
\end{proof}

\subsection{Estimating $m$ (with advice)}

We combine the previously described partial-estimation procedures into two estimation procedures: \procnameZestimateZedgesZadviceZhighHREF, which is optimized for large $m$s, and \procnameZestimateZedgesZadviceZhighHREF, which is optimized for small $m$s.

In \procnameZestimateZedgesZadviceZhighHREF, we use the threshold degree $k = \Theta(\sqrt{n \sqrt{\bar m} / \eps})$, unless $m = \Omega(\eps n^2)$, in which case we use $k = n-1$.

\begin{algo}
    \procname{$\procnameZestimateZedgesZadviceZhigh(\eps, G, \bar m)$}
    \label{alg:e-e-advice-high}
    \alginput{A graph $G=(V,E)$ ($V$ is explicitly given) accessible through \oracleDeg, \oracleN, \oracleIS}
    \alginput{An accuracy parameter $\eps > 0$}
    \alginput{$n \ge 2/\eps^2$}
    \alginput{An untrusted advice $\bar{m} > 0$}
    \algoutput{A number $\tilde \boldm$}
    \begin{code}
        \begin{If} {$\bar m \geq \frac{1}{4}\eps n^2$}
            \algitem Set $k \gets n-1$.
            \algitem Run $\procnameZestimateZllZedgesZadviceHREF(n, \eps, G,\bar m, k)$ and return the result as $\tilde \boldm$.
            \algitem Return $\tilde{\bm}$.
        \end{If}
        \begin{Else}
            \algitem Set $k \gets \frac{\sqrt 2}{\sqrt \eps} n^{\frac{1}{2}}\bar m^{\frac{1}{4}}$.
            \algitem Run $\procnameZestimateZllZedgesZadviceHREF(n, \eps, G,\bar m, k)$ and let $\tilde \boldm_{LL}$ be the result. \label{line:e-e-a-if-2-ell}
            \algitem Run $\procnameZestimateZLoneHZedgesZadviceHREF(n, \eps, G,\bar m, k)$ and let $\tilde \boldm_{L_1H}$ be the result.
            \algitem Return $\tilde \boldm = \tilde \boldm_{LL} + \tilde \boldm_{L_1 H}$.
        \end{Else}
    \end{code}
\end{algo}

\begin{lemma} \label{lm:e-e-advice-high-correctness}
    Let $\tilde{\bm}$ be the (random) output of \procnameZestimateZedgesZadviceZhighHREF (Algorithm \ref{alg:e-e-advice-high}) when given parameters $\eps$, $G = (V,E)$ and $\bar{m}$. If $\bar m \ge \frac{1}{4} m$, then $\abs{\tilde \bm - m} \le 70 \eps \bar m$ with probability at least $1 - \frac{m}{100 \bar m}$.
\end{lemma}
\begin{proof}
    If $\bar{m} \ge \frac{1}{4}\eps n^2$, then we use $k = n-1$. That is, all vertices must have $\deg (u) \le k$, and therefore, $m_{LL} = m$. In this case, by Lemma \ref{lm:e-ll-advice}, $\abs{\tilde\bm - m} \le \eps \bar{m}$ with probability at least $1 - \frac{m}{200\bar{m}}$.
    
    If $\bar{m} < \frac{1}{4}\eps n^2$, then we use $k = \sqrt{2n\sqrt{\bar{m}} / \eps}$.
    \begin{itemize}
        \item By Lemma \ref{lm:e-ll-advice} (the case where $\bar{m} \ge \frac{1}{4}m_{LL}$), with probability at least $1 - \frac{m}{200\bar{m}}$, \procnameZestimateZllZedgesZadviceHREF returns a number $\tilde{\bm}_{LL}$ in the range $m_{LL} \pm \eps \bar{m}$.
        \item By Lemma \ref{lm:e-l1h-advice} (the case where $\bar{m} \ge \frac{1}{4}m_{L_1H}$), with probability at least $1 - \frac{m}{200\bar{m}}$, \procnameZestimateZLoneHZedgesZadviceHREF returns a number $\tilde{\bm}_{L_1H}$ in the range $m_{L_1H} \pm \eps \bar{m}$.
        \item By Lemma \ref{lm:mll-ml1h-cover-most}, there are at most $64 \eps \bar{m}$ missed edges ($(m_{LH} - m_{L_1H}) + m_{HH}$, which is the same as $m - (m_{LL} + m_{L_1H})$).
        \item Combined, $\tilde{\bm}_{LL} + \tilde{\bm}_{L_1H} \in m \pm (1 + 1 + 64)\eps \bar{m} \subseteq m \pm 70 \eps \bar{m}$.
    \end{itemize}

    In both cases ($m \ge \frac{1}{4}\eps^2 n$ and $m < \frac{1}{4}\eps^2 n$), if $\bar{m} \ge \frac{1}{4} m$, then the result is in the range $m \pm 70 \eps \bar{m}$ with probability at least $1 - 2 \cdot \frac{m}{200\bar{m}} = 1 - \frac{m}{100\bar{m}}$.
\end{proof}

\begin{lemma} \label{lm:e-e-advice-high-complexity}
    The expected query complexity of \procnameZestimateZedgesZadviceZhighHREF (Algorithm \ref{alg:e-e-advice-high}) is 
    
    $O\left(\frac{\log n}{\eps^{5/2}} \left(1 + \frac{m}{\bar{m}}\right) \sqrt{n / \sqrt{\bar{m}}} \right)$.
\end{lemma}

\begin{proof}
    If $\bar{m} \ge \frac{1}{4}\eps n^2$, then we run \procnameZestimateZllZedgesZadviceHREF once with $k=n-1$. In this case, we can use $k/\sqrt{\bar m} = O(1/\sqrt{\eps})$, and the expected query complexity is (Lemma \ref{lm:e-ll-advice}):
    \begin{eqnarray*}
        O\left(\frac{\log n}{\eps^2} \cdot \left(1 + \frac{n-1}{\sqrt{\bar m}}\right) \cdot \left(1 + \frac{m}{\mbar}\right)\right)
        = O\left(\frac{\log n}{\eps^2} \cdot \frac{1}{\sqrt{\eps}} \cdot \left(1 + \frac{m}{\mbar}\right)\right)
        = O\left(\frac{\log n}{\eps^{5/2}} \cdot \left(1 + \frac{m}{\mbar}\right) \right).
    \end{eqnarray*}

    If $\bar m < \frac{1}{4}\eps n^2$, then we use $k = \Theta\left(\sqrt{n \sqrt{\bar{m}} /\eps}\right)$. Therefore, the expected complexity of \procnameZestimateZllZedgesZadviceHREF is bounded by (Lemma \ref{lm:e-ll-advice}):
    \begin{eqnarray*}
        O\left(\frac{\log n}{\eps^2} \left(1 + \frac{\sqrt{n \sqrt{\bar{m}} / \eps}}{\sqrt{\bar{m}}}\right) \left(1 + \frac{m}{\bar{m}}\right)\right)
        &=& O\left(\frac{\log n}{\eps^2} \left(1 + \sqrt{n /\eps \sqrt{\bar{m}}}\right) \left(1 + \frac{m}{\bar{m}}\right)\right) \\
        &=& O\left(\frac{\log n}{\eps^{5/2}} \sqrt{n / \sqrt{\bar{m}}} \left(1 + \frac{m}{\bar{m}}\right) \right),
    \end{eqnarray*}

    and the expected complexity of \procnameZestimateZLoneHZedgesZadviceHREF is bounded by (Lemma \ref{lm:e-l1h-advice}):
    \begin{eqnarray*}
        O\left(\frac{1}{\eps^3} \frac{n}{\sqrt{n \sqrt{\bar{m}} /\eps}}\right)
        = O\left(\frac{1}{\eps^{5/2}} \sqrt{n / \sqrt{\bar{m}}}\right).
    \end{eqnarray*}
\end{proof}

\begin{algo}
    \procname{$\procnameZestimateZedgesZadviceZlow(\eps, G, \bar m)$}
    \label{alg:e-e-advice-low}
    \alginput{A graph $G=(V,E)$ ($V$ is explicitly given) accessible through \oracleDeg, \oracleN, \oracleIS}
    \alginput{An accuracy parameter $\eps > 0$}
    \alginput{An untrusted advice $\mbar \ge 1$}
    \algoutput{A number $\tilde \boldm$}
    \begin{code}
        \algitem Set $k \gets \min(\bar m, n - 1)$.
        \begin{If}{$\procnameZsmallZmbarZguardHREF(G,\bar{m})$ accepts}
            \algitem Run $\procnameZestimateZllZedgesZadviceHREF(n, \frac{\eps}{10}, G,\bar m, k)$ and return the result as $\tilde \boldm$.
        \end{If}
        \algitem Return $+\infty$.
    \end{code}
\end{algo}

\begin{lemma} \label{lm:e-e-advice-low-correctness}
    Let $\tilde{\bm}$ be the (random) output of \procnameZestimateZedgesZadviceZlowHREF (Algorithm \ref{alg:e-e-advice-low}) when given parameters $\eps$, $G = (V,E)$ and $\bar{m}$.
    The following is guaranteed for $m = \abs{E}$ (unknown to the algorithm):
    \begin{itemize}
        \item If $\bar m \ge m$, then with probability at least $1 - \frac{m}{200 \bar{m}} - \frac{1}{1000}$, $\abs{\tilde \bm - m} \le \eps \bar m$.
        \item If $\bar m \ge \frac{1}{4}m$, then with probability at least $1 - \frac{m}{200 \bar{m}}$, $\abs{\tilde \bm - m} \le \eps \bar m$ or $\tilde \bm > (1 - \eps)\bar{m}$.
        \item $1 \le \bar m < m$, then with probability at least $1 - \frac{\bar{m}}{200 m}$, $\tilde \bm > (1 - \eps)\bar{m}$.
    \end{itemize}
\end{lemma}
\begin{proof}
    If $\bar{m} \ge m$, then $k = \min(\mbar, n-1) \ge \max_u \deg (u)$, and therefore, $m_{LL} = m$.
    \begin{itemize}
        \item By Lemma \ref{lm:guard-new} (completeness), the guard call accepts with probability at least $1 - \frac{1}{1000}$.
        \item By Lemma \ref{lm:e-ll-advice}, $\abs{\tilde\bm - m} \le \eps \bar{m}$ with probability at least $1 - \frac{m}{200\bar{m}}$.
    \end{itemize}
    The probability that the guard call accepts and the LL-estimation is correct is at least $1 - \frac{m}{200\bar{m}} - \frac{1}{1000}$.

    If $\bar{m} \ge \frac{1}{4}m$, then the guard call may reject (with any probability), in which case we return $+\infty > (1-\eps)\mbar$, and if it does not, then the second bullet still holds.
    
    If $\bar{m} < m$ and $m_{LL} = m$, then by Lemma \ref{lm:e-ll-advice}, $\tilde\bm > (1-\eps)\bar{m}$ with probability at least $1 - \frac{\bar{m}}{200m}$. If the guard call accepts then we correctly return this $\tilde\bm$, and otherwise, we return $+\infty$, which is greater than $(1-\eps)\bar{m}$ as well.

    If $\bar{m} < m$ and $m_{LL} < m$, then there exists a vertex whose degree is greater than $k = \bar{m}$ (note that $k \ne n-1$ since it would imply that $m_{LL}=m$), and therefore (Lemma \ref{lm:guard-new}, soundness case), the guard call rejects with probability at least $1 - \frac{\bar{m}}{1000m}$. In this case, we return $+\infty > (1-\eps)\bar{m}$.
\end{proof}

\begin{lemma} \label{lm:e-e-advice-low-complexity}
    Assume that $\mbar \ge 1$. The expected query complexity of \procnameZestimateZedgesZadviceZlowHREF (Algorithm \ref{alg:e-e-advice-low}) is $O\left(\frac{\log n}{\eps^2} \sqrt{\bar{m}} \right)$.
\end{lemma}
\begin{proof}
    The query complexity of \procnameZsmallZmbarZguardHREF is $O(\sqrt{\bar{m}} \log n)$ (Lemma \ref{lm:guard-new}, complexity).

    If $\bar{m} \ge \frac{1}{4}m$, then by Lemma \ref{lm:e-ll-advice}, the expected query complexity of \procnameZestimateZllZedgesZadviceHREF is
    \[ O\left(\frac{\log n}{\eps^2} \left(1 + \frac{\min(\bar{m},n-1)}{\sqrt{\bar{m}}}\right) \right)
        = O\left(\frac{\log n}{\eps^2} \sqrt{\bar{m}} \right).
    \]

    If $\bar{m} < \frac{1}{4}m$, then by  Lemma \ref{lm:e-ll-advice}, the expected query complexity of \procnameZestimateZllZedgesZadviceHREF is $O\left(\frac{\log n}{\eps^2} \cdot \frac{m}{\bar{m}} \cdot \sqrt{\bar{m}}\right)$. However, by Lemma \ref{lm:guard-new} (soundness), the probability of the guard call to accept is bounded by $O(\bar{m} / m)$, and therefore, the expected query complexity is bounded by
    \[ O(\bar{m} / m) \cdot O\left(\frac{\log n}{\eps^2} \cdot \frac{m}{\bar{m}} \cdot \sqrt{\bar{m}}\right) 
        = O\left(\frac{\log n}{\eps^2} \sqrt{\bar{m}}\right).
    \]

    Combined, the expected query complexity is $O\left(\frac{\log n}{\eps^2} \sqrt{\bar{m}}\right)$.
\end{proof}

\subsection{Estimating $m$ without any advice}

In this subsection we introduce our main procedure, \procnameZestimateZedgesHREF. At a high level, the algorithm begins with an extreme initial estimate for $\bar m$ and iteratively refines it toward an accurate value of $m$. A straightforward serial execution of these iterations (a simple for-loop) achieves correctness with high probability, but its query complexity is small only with high probability rather than in expectation. Therefore, we execute the iterations in an iterative-deepening schedule, i.e., in rounds where we run the prefixes $(0)$, $(0,1)$, $(0,1,2)$, and so on.

We start by presenting a single iteration of the algorithm (Algorithm \ref{alg:e-e-iteration}). We then give an iteration-bounded variant, \procnameZestimateZedgesZboundedHREF (Algorithm \ref{alg:e-e-bounded}), which attains a correct estimate with high probability once the number of iterations is sufficiently large; however, this version guarantees low query complexity only with high probability, not in expectation.

Finally, we provide the iteration-unbounded version, \procnameZestimateZedgesHREF (Algorithm \ref{alg:e-e}), which preserves correctness while ensuring that the expected query complexity remains small.

\begin{algo}
    \procname{$\procnameZestimateZedgesZiteration(n,\eps,G,\ell)$}
    \label{alg:e-e-iteration}
    \alginput{$0 < \eps \le \frac{1}{15}$}
    \alginput{$G$ has $n$ vertices (explicitly given) and $m \ge 1$ edges (unknown to the algorithm)}
    \begin{code}
        \algitem Let $\mbig \gets \frac{n^2}{2^\ell}$, $\msmall \gets 2^{\ell/2}$. \algcomment (\emph{cannot} assume that $\mbig \ge \msmall$)
        \algitem Call $\procnameZestimateZedgesZadviceZhighHREF(n,\frac{1}{1000}\eps,G,\mbig)$ and let $\tildembig$ be the result. \label{line:e-e-a-big}
        \begin{If} {$\frac{1}{4}\mbig \le \tildembig \le \frac{4}{5}\mbig$}
            \algitem Return $\tildembig$ as $\tilde \boldm$. \label{line:e-e-a-return-big}
        \end{If}
        \algitem Call $\procnameZestimateZedgesZadviceZlowHREF(n,\frac{1}{10}\eps,G,\msmall)$ and let $\tildemsmall$ be the result. \label{line:e-e-a-small}
        \begin{If} {$\tildemsmall \leq \frac{1}{\sqrt[4]{2}}\msmall$}
            \algitem Return $\tildemsmall$ as $\tilde \boldm$. \label{line:e-e-a-return-small}
        \end{If}
        \algitem Return \reject.
    \end{code}
\end{algo}

To analyze \procnameZestimateZedgesZiterationHREF (and its callers) when executed on an input graph $G$ with $n$ vertices (where $n$ is known to the algorithm) and $m$ edges (unknown to the algorithm), we define the following parameters:
\begin{itemize}
    \item $\ell_\mathrm{big}(G)$ -- the unique integer for which $\sqrt{2}m \le 2^{-\ell} n^2 < \sqrt{8}m$.
    \item $\ell_\mathrm{small}(G)$ -- the unique integer for which $\sqrt{2}m \le 2^{\ell/2} < 2m$.
    \item $\ell_0(G) = \min(\ell_\mathrm{big}(G), \ell_\mathrm{small}(G))$.
\end{itemize}
For convenience, we usually omit the $G$ parameter and refer to $\ell_\mathrm{big}$, $\ell_\mathrm{small}$ and $\ell_0$.

\begin{lemma} \label{lm:alg-e-e-iteration-correct-small-ell-mbig}
    Assume that $\eps \le \frac{1}{15}$. Considering \procnameZestimateZedgesZiterationHREF (Algorithm \ref{alg:e-e-iteration}), if $\ell \le \ell_0$, then with probability at least $1 - \frac{1}{100} \cdot 2^{(\ell - \ell_0)/2}$, $\tildembig < \frac{1}{4}\mbig(\ell)$ or $\tildembig > \frac{4}{5}\mbig(\ell)$ or $\tildembig \in (1 \pm \eps)m$.
\end{lemma}
\begin{proof}
    By definition, $\ell \le \ell_0$ implies that $\ell \le \ell_\mathrm{big}$ as well. Observe that $\mbig$ is is decreasing monotone in $\ell$. Therefore, $\mbig(\ell) \ge \mbig(\ell_\mathrm{big}) \ge \sqrt{2}m \ge m$.
    
    By Lemma \ref{lm:e-e-advice-high-correctness}, if $\ell \le \ell_\mathrm{big}$, then since $\mbig \ge \frac{1}{4}m$, the failure probability of \procnameZestimateZedgesZadviceZhighHREF is bounded by 
    \[  \frac{m}{100\mbig}
        \le \frac{m}{100 \cdot 2^{-\ell} n^2}
        = \frac{m}{100 \cdot 2^{\ell_\mathrm{big} - \ell} \cdot 2^{-\ell_\mathrm{big}} n^2}
        \le \frac{2^{\ell - \ell_\mathrm{big}} m}{100 \cdot \sqrt{2}m}
        \le \frac{2^{\ell - \ell_0}}{100\sqrt{2}}
        \le \frac{1}{100} \cdot 2^{(\ell - \ell_0) / 2}.
    \]
    
    Note that $70 \cdot \frac{1}{1000}\eps < \frac{1}{12}\eps$. Conditioned on not failing, $\tildembig \in m \pm \frac{1}{12}\mbar$, and we can use the following case analysis:
    \begin{itemize}
        \item If $\ell \le \ell_\mathrm{big} - 2$, then $\bar m = \mbig(\ell)=2^{-\ell}n^2\ge 4\cdot 2^{-\ell_{\mathrm{big}}}n^2\ge 4\sqrt{2}m$, and therefore,
        $\tildembig \le m+\frac{\eps}{12}\bar m \le  \frac{1}{4\sqrt{2}}\bar{m} + \frac{1}{12}\eps \bar{m} < \frac{1}{4}\bar{m}$.
        \item If $\ell = \ell_\mathrm{big} - 1$, then $\bar{m} < 2 \cdot \sqrt{8} m < 12m$ and hence $\tildembig \in (1 \pm \eps)m$.
        \item If $\ell = \ell_\mathrm{big}$, then $\bar{m} < 4 \cdot \sqrt{8} m < 12m$ and hence $\tildembig \in (1 \pm \eps)m$.
    \end{itemize}
\end{proof}

\begin{lemma} \label{lm:alg-e-e-iteration-correct-small-ell-msmall}
    Assume that $\eps \le \frac{1}{10}$. Considering \procnameZestimateZedgesZiterationHREF (Algorithm \ref{alg:e-e-iteration}), if $\ell \le \ell_0 - 2$, then with probability at least $1 - \frac{1}{100} \cdot 2^{(\ell - \ell_0)/2}$, $\tildemsmall > \frac{1}{\sqrt[4]{2}}\msmall$ or $\tildemsmall \in (1 \pm \eps)m$. If $\ell = \ell_0 - 1$, then the probability is only at least $1 - \frac{1}{100}$.
\end{lemma}
\begin{proof}
    Observe that $\ell \le \ell_0 - 1$ implies that $\ell \le \ell_\mathrm{small} - 1$. We have two cases: $\ell \le \ell_\mathrm{small} - 2$ (this implies that $\ell \le \ellsmall - 2$) and $\ell = \ell_\mathrm{small} - 1$.

    Case I: $\ell \le \ell_\mathrm{small} - 2$. Observe that $\msmall(\ell) \le \msmall(\ell_\mathrm{small} - 2) = 2^{-2/2} \msmall(\ell_\mathrm{small}) < \frac{1}{2} \cdot 2m = m$. By Lemma \ref{lm:e-e-advice-low-correctness}, the probability that $\tilde\bm \le \left(1-\frac{1}{10}\eps\right)\msmall$ (which contains the bad event $\tilde\bm \le \frac{1}{\sqrt[4]{2}}\msmall$) is bounded by:
    \[  \frac{\msmall}{200 m}
        = \frac{2^{(\ell-\ellsmall)/2} \msmall(\ellsmall)}{200 m}
        \le \frac{2^{(\ell-\ellsmall)/2} \cdot 2m}{200 m}
        = \frac{1}{100} \cdot 2^{(\ell-\ellsmall)/2}.
    \]
    
    Case II. $\ell = \ellsmall - 1$. Note that
    \[  \frac{1}{4}m
        < \frac{1}{\sqrt{2}}m
        \le 2^{-2/2}\msmall(\ellsmall)
        \le \msmall(\ell)
        = 2^{-1/2}\msmall(\ellsmall)
        < \sqrt{2} m.
    \]

    By Lemma \ref{lm:e-e-advice-low-correctness} (the case where $\bar{m} \ge \frac{1}{4} m$), $\tildemsmall > (1 - \eps)\msmall > \frac{1}{\sqrt[4]{2}}\msmall$ or $\abs{\tildemsmall - m} \le \frac{1}{10}\eps \msmall \le \eps m$ with probability at least $1 - \frac{\msmall}{200m} \ge 1 - \frac{1}{100}$.
\end{proof}

\begin{lemma} \label{lm:alg-e-e-iteration-correct-small-ell}
    Assume that $\eps \le \frac{1}{15}$. If $\ell \le \ell_0 - 2$, then with probability at least $1 - \frac{1}{50} \cdot 2^{(\ell-\ell_0)/2}$, \procnameZestimateZedgesZiterationHREF (Algorithm \ref{alg:e-e-iteration}) either rejects or returns a number in the range $(1 \pm \eps) m$. If $\ell = \ell_0 - 1$, then this probability is at least $1 - \frac{1}{100} \cdot 2^{(\ell-\ell_0)/2} - \frac{1}{100}$.
\end{lemma}
\begin{proof}
    Just union bound Lemma \ref{lm:alg-e-e-iteration-correct-small-ell-mbig} and Lemma \ref{lm:alg-e-e-iteration-correct-small-ell-msmall}.
\end{proof}

\begin{lemma} \label{lm:alg-e-e-iteration-correct-exact-ell0}
    For $\eps \le \frac{1}{15}$, if $\ell = \ell_0$, then \procnameZestimateZedgesZiterationHREF (Algorithm \ref{alg:e-e-iteration}) returns a number in the range $(1 \pm \eps) m$ with probability at least $1 - \frac{1}{55}$.
\end{lemma}
\begin{proof}
    We have two cases: $\ell_0 = \ell_\mathrm{big}$ and $\ell_0 = \ell_\mathrm{small}$.

    Case I: $\ell_0 = \ell_\mathrm{big}$. In this case, $\sqrt{2}m \le \mbig < \sqrt{8} m$, and therefore, with probability at least $1 - \frac{m}{100 \mbig} \ge 1 - \frac{1}{100\sqrt{2}}$,
    $\tildembig \in m \pm \frac{1}{10}\mbig \subseteq (1 \pm \frac{\sqrt{8}}{10}m) \subseteq (1 \pm \eps)m$ (Lemma \ref{lm:e-e-advice-high-correctness}). Also,
    \[\frac{1}{4}\mbig \le \frac{1}{\sqrt{2}}m \le (1 - \eps)m \le \tildembig \le (1 + \eps)m \le \frac{1+\eps}{\sqrt{2}}\mbig \le \frac{4}{5}\mbig,\]
    and hence, we hit line \ref{line:e-e-a-big} and return $\tildembig$.

    Case II: $\ell_0 = \ell_\mathrm{small}$ (and $\ell_0 \le \ell_\mathrm{big} - 1$). The bad event is:
    \begin{itemize}
        \item $\tildembig \ge \frac{1}{4}\mbig$ and $\tildembig \le \frac{4}{5}\mbig$ and $\tildembig \notin (1 \pm \eps)m$, or
        \item $\tildemsmall \notin (1 \pm \eps)m$.
    \end{itemize}

    By Lemma \ref{lm:alg-e-e-iteration-correct-small-ell-mbig}, the probability of the first sub-event is bounded by $\frac{1}{100} \cdot 2^{(\ell - \ell_0)/2} = \frac{1}{100}$.

    By Lemma \ref{lm:e-e-advice-low-correctness} (the $\bar{m} \ge m$ case), since $\msmall(\ell) = \msmall(\ell_0) \ge \sqrt{2}m$, the probability of the second sub-event is bounded by $\frac{m}{200\mbar} + \frac{1}{1000} \le \frac{1}{100\sqrt{2}} + \frac{1}{1000}$.
    
    By the union bound, the probability of the bad event in the case where $\ell_0 = \ell_\mathrm{small}$ is at most $\frac{1}{100} + \frac{1}{100\sqrt{2}} + \frac{1}{1000} < \frac{1}{55}$.
\end{proof}

\begin{lemma} \label{lm:e-e-complexity-specific-iteration-new-part-i}
    The expected query complexity of obtaining $\tildembig$ in \procnameZestimateZedgesZiterationHREF is $O(2^{\max(0, \ell-\ell_0)} \cdot 2^{\ell/4} \log n / \eps^{5/2})$.
\end{lemma}
\begin{proof}
    By Lemma \ref{lm:e-e-advice-high-complexity}, the query complexity of \procnameZestimateZedgesZadviceZhighHREF is 
    \begin{eqnarray*}
        O\left(\frac{\log n}{\eps^{5/2}} \left(1 + \frac{m}{\mbig}\right) \sqrt{n / \sqrt{\mbig}}\right)
        &=& O\left(\frac{\log n}{\eps^{5/2}} \left(1 + \frac{m}{\mbig}\right) \sqrt{n / \sqrt{2^{-\ell} n^2}}\right) \\
        &=& O\left(\frac{\log n}{\eps^{5/2}} \left(1 + \frac{m}{\mbig}\right) 2^{\ell/4}\right).
    \end{eqnarray*}

    If $\ell \le \ellbig$, then $\mbig(\ell) \ge \mbig(\ellbig) \ge \sqrt{2}m$, and therefore, $O\left(1 + \frac{m}{\mbig}\right) = O(1)$.

    If $\ell > \ellbig$, then
    \begin{eqnarray*}
        O\left(1 + \frac{m}{\mbig}\right)
        = O\left(\frac{m}{\mbig}\right)
        &=& O\left(\frac{m}{2^{-(\ell-\ellbig)} \mbig(\ellbig)}\right) \\
        &=& O\left(2^{\ell-\ellbig} \frac{m}{\sqrt{2} m}\right)
        = O\left(2^{\ell-\ellbig}\right)
        = O\left(2^{\ell-\ell_0}\right).
    \end{eqnarray*}

    Combined, the expected query complexity is $O\left(2^{\max(\ell-\ell_0,0)} \cdot \frac{\log n}{\eps^{5/2}} 2^{\ell/4}\right)$.
\end{proof}

\begin{lemma} \label{lm:e-e-complexity-specific-iteration-new-part-ii}
    The expected query complexity of obtaining $\tildemsmall$ in \procnameZestimateZedgesZiterationHREF is $O(2^{\ell/4} \log n / \eps^{5/2})$.
\end{lemma}
\begin{proof}
    By Lemma \ref{lm:e-e-advice-low-complexity}, the query complexity of \procnameZestimateZedgesZadviceZhighHREF is $O\left(\frac{\log n}{\eps^2} \sqrt{\msmall}\right)$, which is $O\left(\frac{\log n}{\eps^2} \cdot 2^{\ell/4}\right)$.
\end{proof}

\begin{lemma} \label{lm:e-e-complexity-specific-iteration}
    The expected query complexity of \procnameZestimateZedgesZiterationHREF is $O(2^{\max(0, \ell-\ell_0)} \cdot 2^{\ell/4} \cdot \log n / \eps^{5/2})$.
\end{lemma}
\begin{proof}
    Just combine Lemma \ref{lm:e-e-complexity-specific-iteration-new-part-i} and Lemma \ref{lm:e-e-complexity-specific-iteration-new-part-ii}.
\end{proof}

\begin{algo}
    \procname{$\procnameZestimateZedgesZbounded(n,\eps,G,\ell_\mathrm{max})$}
    \label{alg:e-e-bounded}
    \alginput{$0 < \eps \le \frac{1}{10}$}
    \alginput{$G$ has $n$ vertices (explicitly given) and $m$ edges (unknown to the algorithm)}
    \alginput{$m \ge 1$}
    \begin{code}
        \begin{For}{$\ell$ from $0$ to $\ell_\mathrm{max}$}
            \algitem Cal $\procnameZestimateZedgesZiterationHREF(n,\eps,G,\ell)$ and let $\tilde \bm$ be its result.
            \begin{If}{$\tilde \bm \ne \reject$}
                \algitem Return $\tilde \bm$.
            \end{If}
        \end{For}
        \algitem Return \reject.
    \end{code}
\end{algo}

\begin{lemma} \label{lm:alg-e-e-bounded-correctness}
    Assume that $\eps \le \frac{1}{15}$. Considering \procnameZestimateZedgesZboundedHREF:
    \begin{itemize}
        \item If $\ell_\mathrm{max} \le \ell_0 - 2$, then the probability of the $\ell$th call to \procnameZestimateZedgesZiterationHREF to return a non-reject value outside the range $(1 \pm \eps)m$ is bounded by $\frac{2+\sqrt{2}}{50} \cdot 2^{(\ell - \ell_0) / 2}$.
        \item If $\ell_\mathrm{max} = \ell_0 - 1$, then the probability to return a number outside the range $(1 \pm \eps)m$ is bounded by $\frac{3}{50}$.
        \item If $\ell_\mathrm{max} \ge \ell_0$, then the probability to return a number in the range $(1 \pm \eps)m$ is at least $1 - 3/34$.
    \end{itemize}
\end{lemma}
\begin{proof}
    Case I. $\ell_\mathrm{max} \le \ell_0 - 2$. By Lemma \ref{lm:alg-e-e-iteration-correct-small-ell} and by the union bound, the probability to return a number outside the range $(1 \pm \eps)m$ is at most:
    \[  \sum_{\ell=0}^{\ell_\mathrm{max}} \frac{1}{50} \cdot 2^{(\ell-\ell_0)/2}
        = \frac{1}{50} \cdot 2^{-\ell_0/2} \sum_{\ell=0}^{\ell_\mathrm{max}} 2^{\ell/2}
        \le \frac{1}{50} \cdot 2^{-\ell_0 / 2} \cdot \left((2 + \sqrt{2}) \cdot 2^{\ell_\mathrm{max} / 2}\right)
        = \frac{2 + \sqrt{2}}{50} \cdot 2^{(\ell_\mathrm{max} - \ell_0) / 2}.
    \]

    For $\ell_\mathrm{max} = \ell_0 - 1$,
    \begin{eqnarray*}
        \sum_{\ell=0}^{\ell_0 - 2} \frac{1}{50} \cdot 2^{(\ell-\ell_0)/2} + \left(\frac{1}{100} \cdot 2^{-1/2} + \frac{1}{100}\right)
        &\le& \frac{2+\sqrt{2}}{50}2^{-2/2} + \frac{1 + 1/\sqrt{2}}{100} \\
        &=& \frac{2+\sqrt{2} + 1 + 1/\sqrt{2}}{100}
        \le \frac{3}{50}.
    \end{eqnarray*}

    For $\ell \ge \ell_0$, note that by monotonicity, it suffices to bound the fail probability for $\ell = \ell_0$, which is bounded by:
    \[  \underbrace{3 / 50}_{\text{bound for $\ell_0-1$}} + \underbrace{1 / 55}_{\text{individual fail probability of $\ell_0$}}
        < \frac{3}{34}.
    \]
\end{proof}

\begin{lemma} \label{lm:alg-e-e-bounded-complexity}
    The expected query complexity of \procnameZestimateZedgesZboundedHREF is  $O(2^{\max(\ell_\mathrm{max} - \ell_0,0)} \cdot 2^{\ell/4} \log n / \eps^{5/2})$.
\end{lemma}
\begin{proof}
    By Lemma \ref{lm:e-e-complexity-specific-iteration} and by linearity of expectation (sum over the individual iterations), the expected query complexity is bounded by:
    \begin{eqnarray*}
        \sum_{\ell=0}^{\ell_\mathrm{max}} O(2^{\max(\ell - \ell_0, 0)} 2^{\ell/4} \log n / \eps^{5/2})
        &=& O(\log n / \eps^{5/2}) \cdot \sum_{\ell=0}^{\ell_\mathrm{max}} 2^{\ell/4} 2^{\max(\ell-\ell_0,0)} \\
        &=& O(\log n / \eps^{5/2}) \cdot O(1) \cdot 2^{\ell_\mathrm{max}/4 + \max(\ell_\mathrm{max} - \ell_0,0)} \\
        &=& O(2^{\max(\ell_\mathrm{max} - \ell_0,0)} \cdot 2^{\ell/4} \log n / \eps^{5/2}).
    \end{eqnarray*}
\end{proof}

\begin{algo}
\procname{$\procnameZestimateZedges(n,\eps,G)$}
\label{alg:e-e}
    \alginput{Integer $n$, parameter $\eps \in (0, 1]$}
    \alginput{Access to the oracles \oracleDeg, \oracleN, \oracleIS of a graph $G=([n], E)$}
    \algoutput{A number $\tilde \boldm$, an estimate for $m$}
    \begin{code}
        \begin{If} {$\eps > \frac{1}{15}$}
            \algitem Set $\eps \gets \frac{1}{15}$.
        \end{If}
        \begin{If} {$\oracleIS(V) = 1$}
            \algitem Return $0$.
        \end{If}
        \begin{If}{$n < 2/\eps^2$}
            \begin{For}{each $u \in V$}
                \algitem Let $d_u \gets \deg (u)$ (via \oracleDeg).
            \end{For}
            \algitem Return $\frac{1}{2}\sum_{u \in V} d_u$. \label{line:e-e-smalln-case}
        \end{If}
        \algitem Let $\ell \gets 0$.
        \begin{While}{\textsc{True}}
            \algitem Call $\procnameZestimateZedgesZboundedHREF(n,\eps,G,\ell)$ and let $\tilde \bm$ be its result.
            \begin{If}{$\tilde\bm \ne \reject$}
                \algitem Return $\tilde\bm$.
            \end{If}
            \algitem $\ell\leftarrow \ell+1$.
        \end{While}
    \end{code}
\end{algo}

\begin{lemma} \label{lm:e-e-correctness}
    Let $\eps \le \frac{1}{15}$ and $G$ be the input graph. With probability at least $2/3$, Algorithm \ref{alg:e-e} returns an estimation in the range $(1 \pm \eps)m$.
\end{lemma}
\begin{proof}
    By Lemma \ref{lm:alg-e-e-bounded-correctness}, the probability to return a non-reject value outside the range $(1 \pm \eps)m$ within the first $\ell_0-1$ iterations (each consisting of a call to \procnameZestimateZedgesZboundedHREF) is bounded by:
    \begin{eqnarray*}
        \sum_{\ell=0}^{\ell_0 - 2} \frac{2+\sqrt{2}}{50} 2^{(\ell - \ell_0) / 2} + \frac{3}{50}
        \le \frac{(2 + \sqrt{2})^2}{50} \cdot 2^{-2/2} + \frac{3}{50}
        \le 0.18.
    \end{eqnarray*}

    If we execute the $\ell_0$th iteration, then by Lemma \ref{lm:alg-e-e-bounded-correctness}, the probability to return a number in the range $(1 \pm \eps)m$ is at least $1 - 3/34$.

    Combined, the probability to return a number in the range $(1 \pm \eps)m$ is at least $\left(1 - 0.18\right) \cdot \left(1 - \frac{3}{34}\right) > \frac{2}{3}$.
\end{proof}

\begin{lemma} \label{lm:e-e-complexity}
    For $\eps \le \frac{1}{15}$ and an input graph $G$ with $n$ vertices and $m$ edges, the expected query complexity of Algorithm \ref{alg:e-e} is bounded by $O\left(\frac{\log n}{\eps^{5/2}} \min\left(\sqrt{m}, \sqrt{\frac{n}{\sqrt{m}}}\right)\right)$.
\end{lemma}
\begin{proof}
    For every $\ell \ge 0$, let $T_\ell$ be the expected query complexity of the $\ell$th iteration. By Lemma \ref{lm:alg-e-e-bounded-complexity}, there exists some $C = O(\log n / \eps^{5/2})$ for which $T_\ell \le C \cdot 2^{\max(0, \ell - \ell_0)} \cdot 2^{\ell/4}$ for every $\ell \ge 0$.
    
    By Lemma \ref{lm:alg-e-e-bounded-correctness}, for every $\ell \ge \ell_0$, the probability to reject in the $\ell$th call to \procnameZestimateZedgesZboundedHREF is at most $3/34 < 1/6$. Therefore, the expected query complexity of \procnameZestimateZedgesHREF is bounded by:
    \begin{eqnarray*}
        \sum_{\ell=0}^{\ell_0} T_\ell + \sum_{\ell=\ell_0+1}^\infty \frac{1}{6^{\ell-\ell_0}} T_\ell
        &\le& C \sum_{\ell=0}^{\ell_0} 2^{\ell/4} + C \sum_{\ell=\ell_0+1}^\infty 2^{\ell-\ell_0} \cdot 2^{\ell/4} \cdot 6^{\ell_0-\ell} \\
        &=& C \cdot 2^{\ell_0/4} \sum_{\ell=0}^{\ell_0} 2^{(\ell-\ell_0)/4} + C \sum_{\ell=\ell_0+1}^\infty 2^{\ell-\ell_0} \cdot 2^{\ell_0/4} \cdot 2^{(\ell-\ell_0)/4} \cdot 6^{\ell_0-\ell} \\
        &\le& C \cdot 2^{\ell_0/4} \left(\sum_{i=0}^\infty 2^{-i/4} + \sum_{i=1}^\infty (2^{5/4} / 6)^i\right) \\
        &\le& C \cdot 2^{\ell_0/4} (6.29 + 0.66)
        \le 7 C \cdot 2^{\ell_0/4}.
    \end{eqnarray*}

    To complete the proof, we have to show that $2^{\ell_0/4} = O(\min(\sqrt{m}, \sqrt{n / \sqrt{m}}))$:
    \begin{eqnarray*}
        2^{\ell_0/4}
        = \min\left(2^{\ellsmall/4}, 2^{\ellbig / 4} \right)
        &=& \min\left(\sqrt{2^{\ellsmall/2}}, \sqrt{n / \sqrt{2^{-\ellbig} n^2}} \right) \\
        &=& \min\left(\sqrt{\msmall(\ellsmall)}, \sqrt{n / \sqrt{\mbig(\ellbig)}} \right) \\
        &=& \min\left(\sqrt{\Theta(m)}, \sqrt{n / \sqrt{\Theta(m)}} \right)
        = O\left(\min\left(\sqrt{m}, \sqrt{n / \sqrt{m}} \right)\right).
    \end{eqnarray*}
\end{proof}

\begin{lemma} \label{lem:ubnd}
    For $\eps \le \frac{1}{15}$ and an input graph $G$ with $n$ vertices and $m$ edges. \procnameZestimateZedgesHREF (Algorithm \ref{alg:e-e}) returns a number in the range $(1 \pm \eps)m$ with probability at least $2/3$. Moreover, 
    its expected query complexity bounded by $O\left(\frac{\log n}{\eps^{5/2}} \min\left(\sqrt{m}, \sqrt{\frac{n}{\sqrt{m}}}
    \right)\right)$.
\end{lemma}
\begin{proof} The proof follows from
    Lemma \ref{lm:e-e-correctness} (correctness) and Lemma \ref{lm:e-e-complexity} (complexity).
\end{proof}

\section{Lower Bound}

In this section we prove the following lower-bound for edge estimation in the hybrid model. In particular, the lower bound follows from the following (stronger) statement.
\begin{theorem} \label{th:lower-bound-main}
    Consider any algorithm that is given query access to an unknown graph $G=(V,E)$ on $n$ vertices and $m$ edges via degree queries, neighbor queries and IS queries. Let $\eps\in(0,1)$ and define $R\eqdef\frac{1}{30}\min\left(\sqrt m, \sqrt{\frac{n}{\eps\sqrt{m}}}\right)$.
    For any such algorithm whose expected number of queries is $q$ (where $q$ may depend on $G$), the probability that the algorithm outputs an estimate $\widehat{\bm}$ satisfying $(1-\eps)m\le \widehat \bm\le (1+\eps)m$ is at most $\frac{1}{2} + q/R$.
\end{theorem}

Theorem~\ref{th:lower-bound-main} above bounds the success probability of any algorithm as a function of its expected number of queries. Rearranging this bound and applying Markov's inequality yields the following corollary for algorithms that succeed with probability at least  $2/3$.
\begin{corollary}
    Let $\eps\in(0,1)$. Every algorithm whose input is an access to a graph $G$ of $m$ edges (where $m$ is unknown to the algorithm) through degree queries,  neighbor queries and IS queries, whose output $\widehat{\bm}$ satisfies  $(1-\eps)m\le \widehat \bm\le (1+\eps)m$ with probability at least $2/3$, must make $\Omega\left(\min\left(\sqrt{m}, \sqrt{\frac{n}{\eps\sqrt m}}\right)\right)$ queries.
\end{corollary}
The proof of Theorem~\ref{th:lower-bound-main} proceeds via a reduction from a specially designed string-testing problem to the problem of estimating the number of edges in a graph with a particular structure.

\subsection{The string distinction problem}

Let $s$ be a string of length $n$ over the alphabet $\Sigma = \{ \mathrm{A}, \mathrm{K}, \mathrm{L}, \mathrm{H} \}$. We define a duality map as follows: $\mathrm{A}$ and $\mathrm{K}$ are the dual of themselves ($\overline{\mathrm{A}} = \mathrm{A}$, $\overline{\mathrm{K}} = \mathrm{K}$), and $\mathrm{L}$ and $\mathrm{H}$ are the dual of each other ($\overline{\mathrm{L}} = \mathrm{H}$, $\overline{\mathrm{H}} = \mathrm{L}$).

For $\sigma \in \Sigma$ and a string $s \in \Sigma^n$, let $n_\sigma(s) = \abs{\{1 \le i \le n : s(i) = \sigma\}}$. For every $1 \le i \le n$, let
\[
U_i \eqdef
\begin{cases}
\emptyset, & \text{if } s(i) = \mathrm{A},\\[4pt]
\{\, j \in [n]\setminus\{i\} : s(j) = \overline{s(i)} \,\}, & \text{otherwise}.
\end{cases}
\]

We assume that $s$ is accessible through two oracles:
\begin{itemize}
    \item \textbf{Full query.} Given an index $1 \le i \le n$, the oracle returns three entries:
    \begin{itemize}
        \item The symbol $s(i)$.
        \item The value $\abs{U_i}$. 
        \item A uniformly random index from $U_i$, or $\bot$ if $U_i=\emptyset$.
    \end{itemize}
    \item \textbf{Light query.} Given an index $1 \le i \le n$, the oracle returns the symbol $s(i)$.
\end{itemize}

We consider the task of distinguishing between strings of length $n$ for which $n_\mathrm{L}(s) \cdot n_\mathrm{H}(s) = 0$ and strings for which $n_\mathrm{L}(s) \cdot n_\mathrm{H}(s) > \eps \binom{n_\mathrm{K}(s)}{2}$.

We say that an algorithm \emph{distinguishes} between the above cases if one of the following occurs:
\begin{definition}[Full-query distinguishing predicate]
    An algorithm performs a full query on a vertex $i$ for which $s(i) \in \{ \mathrm{L}, \mathrm{H} \}$ and obtained a non-$\bot$ index as the third entry of the result tuple.
\end{definition}

\begin{definition}[Light-query distinguishing predicate]
    An algorithm performs a light query on a vertex $i$ for which $s(i) = \mathrm{H}$ and another light query on a vertex $j$ for which $s(j) = \mathrm{L}$.
\end{definition}

Observe that if either predicate occurs, the algorithm can deduce that $n_{\mathrm{L}}(s) \cdot n_{\mathrm{H}}(s) \ne 0$. Conversely, if neither predicate occurs, the algorithm cannot distinguish between the case where all $\{\mathrm{L},\mathrm{H}\}$ symbols are the same (all $\mathrm{L}$ or all $\mathrm{H}$) and the case where they are not.

\subsection{Reduction to edge estimation}

\begin{figure}[htp]
    \centering
    \includegraphics[width=12cm]{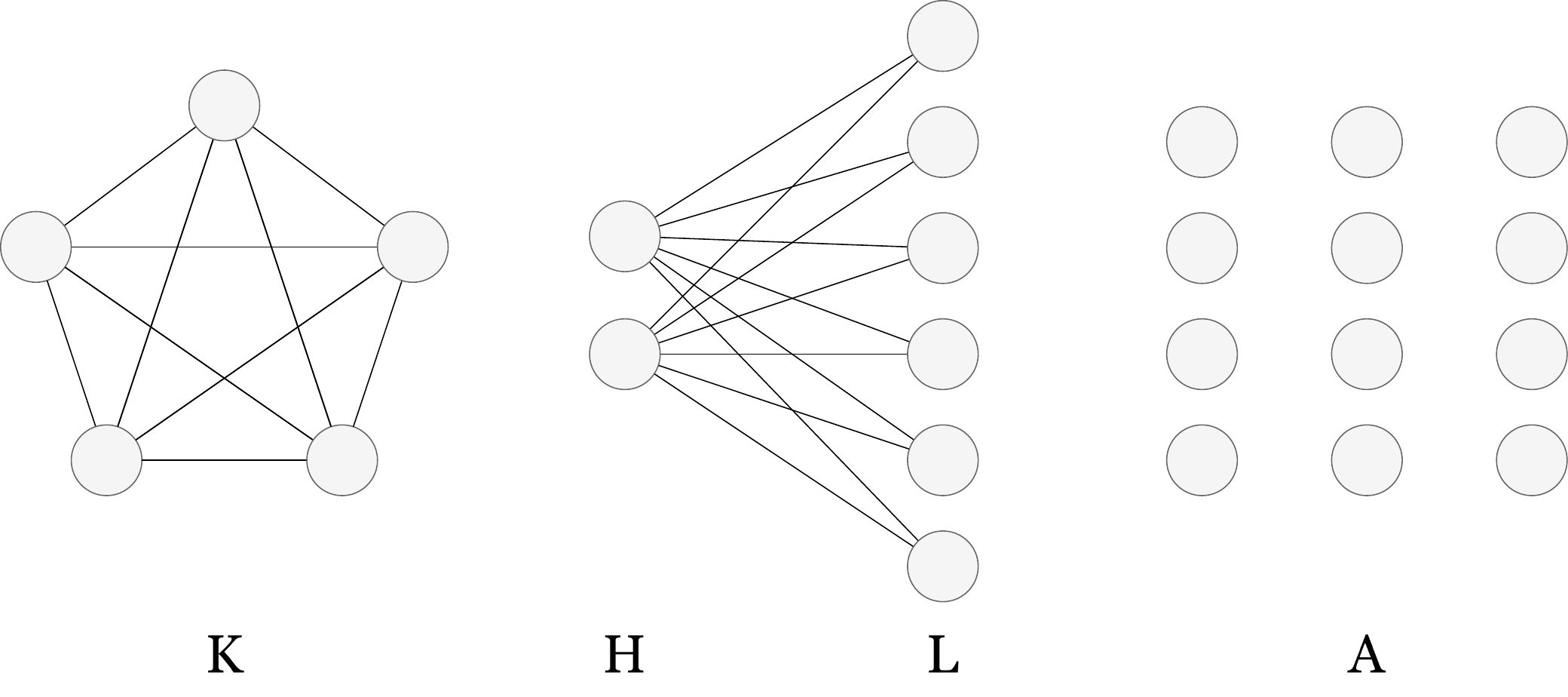}
    \caption{A diagram of the graph structure}
    \label{fig:svg:clique-and-biclique}
\end{figure}

We can interpret a string $s$ as a description of a graph of the following structure. A clique over the set $\{ i : s(i) = \mathrm{K} \}$ and a biclique whose sides are $\{i : s(i) = \mathrm{L} \}$ and $\{i : s(i) = \mathrm{H} \}$. The number of edges is exactly $\binom{n_\mathrm{K}(s)}{2} + n_\mathrm{L}(s) \cdot n_\mathrm{H}(s)$. More formally,

\begin{definition}[Reduction graph]
    Let $s \in \Sigma^n$ be a string. The \emph{reduction graph} $G_s$ is the graph with vertex set $\{1,\ldots,n\}$ defined as follows. An edge is placed between every pair of vertices $i,j$ such that $s(i)=s(j)=\mathrm{K}$, and between every pair of vertices $i,j$ such that $s(i)=\mathrm{L}$ and $s(j)=\mathrm{H}$. Equivalently, $G_s$ consists of a clique induced by the vertices labeled $\mathrm{K}$ and a complete bipartite graph between the vertices labeled $\mathrm{L}
    $ and those labeled $\mathrm{H}$.  Figure~\ref{fig:svg:clique-and-biclique} illustrates the structure of $G_s$.
\end{definition}

We describe a reduction that simulates graph queries using string queries. At the $t$-th step of the simulation, the following process is executed:
\begin{enumerate}
    \item The $t$-th graph query, denoted $\Lambda_t$, is determined as a function of the outcomes of the preceding graph queries and the randomness of the algorithm.
    \item The query $\Lambda_t$ is translated into a sequence of string queries to the string $s$. 
    Let $I_{t,1},\ldots,I_{t,M_t}$ denote this sequence, where $M_t$ is the number of string queries performed at step $t$. 
    The value of $M_t$ may depend on the outcomes of previous string queries, on the string $s$, and on the internal randomness of the translation algorithm.
    \item Using the responses to the string queries $I_{t,1},\ldots,I_{t,M_t}$, the simulation deterministically computes $Y_t$, the outcome of the $t$-th graph query.
\end{enumerate}

At this point we describe the translation algorithm. Degree and neighbor queries are simulated using a single full string query, whereas an independent-set query is simulated by a randomized, adaptively determined sequence of light string queries.

Formally, the simulation proceeds as follows.
\begin{itemize}
    \item \textbf{Degree query.} Given a vertex $u$ we issue a full string query to $s$ at position $u$ and return the second component of the resulting triplet.
    \item \textbf{Neighbor query.} Given a vertex $u$ we issue a full string query to $s$ at position $u$ and return the third component of the resulting triplet.
    \item \textbf{Independent-set query.} Given a vertex set $S$ we first apply a uniformly random permutation to the vertices of $S$. We then process the vertices in this order, issuing light string queries, until one of the following stopping conditions is met:
    \begin{enumerate}
        \item Two vertices labeled $\mathrm{K}$ are revealed, certifying that $S$ is not an independent set.
        \item One vertex labeled $\mathrm{L}$ and one vertex labeled $\mathrm{H}$ are revealed, certifying that $S$ is not an independent set.
        \item All vertices in $S$ have been processed without triggering either of the above conditions, in which case $S$ is declared an independent set.
    \end{enumerate}

\end{itemize}
\begin{remark}\label{rem:only-unseen}
    In the simulation of an independent-set query, whenever the symbol of a vertex has already been revealed, either by a previous string query or as part of the output of another query, we reuse this information and do not issue an additional query for that vertex.
\end{remark}

The correctness of the simulation is trivial by the construction details. In the following we bound the simulation complexity. For local queries this is directly implied from the implementation:
\begin{observation}
    The simulation of any local query (degree or neighbor) incurs exactly one full string query.
\end{observation}

It remains to bound the expected number of light string queries used in the simulation of an independent-set query.

\begin{lemma} \label{lm:lbnd:graph-query-involve-2k}
A vertex is said to be \emph{involved} in a string query if it is either (i) directly queried, or (ii) appears in the output of that query. Every simulated graph query involves at most two vertices $u$ such that $s(u)=\mathrm{K}$.
\end{lemma}
\begin{proof}
We consider the two types of simulated graph queries.

For a local query (degree or neighbor), the simulation performs a single full string query. Such a query involves the queried vertex and at most one additional vertex appearing as part of the output (as the third entry). Hence, the simulation of a local query involves at most two vertices in total, and therefore at most two vertices labeled $\mathrm{K}$.

For an independent-set query, the simulation issues a sequence of light string queries, each involving exactly one vertex (the queried vertex). By construction, the simulation halts immediately once it has revealed two vertices $u$ with $s(u)=\mathrm{K}$. Consequently, throughout the execution of the simulation, at most two $\mathrm{K}$-vertices can be involved in the string query.
\end{proof}

\begin{lemma} \label{lm:lbnd:IS-translation-string-queries}
Assume that $\bs \in \{\mathrm{K},\mathrm{A}\}^n$ is drawn from a distribution that is invariant under permutations of its coordinates; that is, all permutations of $\bs$ have equal probability (while strings that are not permutations of one another may have different probabilities). Suppose that the $t$-th query issued by the graph algorithm is an independent-set query. 
If $t < \tfrac{1}{2} n_{\mathrm{K}}(\bs)$, then conditioned on any outcomes of the first $t-1$ queries, the simulation of the $t$-th query performs at most
$\frac{2n}{n_{\mathrm{K}}(\bs)-2t}$
light string queries in expectation.
\end{lemma}
\begin{proof}
    By Lemma \ref{lm:lbnd:graph-query-involve-2k}, each of the previous $t-1$ queries involves at most two $\mathrm{K}$-vertices, which is at most $2(t-1)$ in total. Consequently, conditioned on the outcomes of the first $t-1$ queries, when the simulation queries a vertex $u$ that has not previously been involved in any string query, the conditional probability that it satisfies $\bs(u) = \mathrm{K}$ is at least $\frac{n_\mathrm{K}(\bs)-2(t-1)}{n}$.
    
    Let $S$ be the vertex set queried in the $t$-th iteration, and let $S_{\mathrm{unseen}} \subseteq S$ denote the subset of vertices whose symbols have not been revealed prior to this iteration, namely, vertices that have neither been directly queried nor appeared as part of the output of any previous string query.

    By the simulation rule described above (see Remark~\ref{rem:only-unseen}), the translation of the $t$-th graph query issues string queries only to vertices in $S_{\mathrm{unseen}}$: if a vertex in $S$ has already been revealed, its symbol is reused and no additional string query is performed. Consequently, during the simulation of the $t$-th query, newly revealed vertices can only come from $S_{\mathrm{unseen}}$.

    Before issuing the $t$-th query, any vertex $u \in S_{\mathrm{unseen}}$ has conditional probability at least $\frac{n_{\mathrm{K}}(\bs)-2(t-1)}{n}$ of satisfying $\bs(u) = \mathrm{K}$, conditioned on the outcomes of the first $t-1$ queries. Therefore, the number of string queries performed until a $\mathrm{K}$-vertex is revealed is dominated by a geometric random variable  whose expected value is $\frac{n}{n_{\mathrm{K}}(\bs)-2(t-1)} \le \frac{n}{n_{\mathrm{K}}(\bs)-2t}.$

    If a $\mathrm{K}$-vertex is found, then the probability of a still unseen vertex to have $s(u) = \mathrm{K}$ is at least $\frac{n_\mathrm{K}(s) - 2(t-1) - 1}{ n}$, and therefore, the number of queries performed until a second $\mathrm{K}$-vertex is found is bounded by a geometric variable whose expected value is $\frac{n} { n_\mathrm{K}(s) - 2(t-1) - 1} \le \frac{n}{n_\mathrm{K}(s) - 2t}$ as well.

    Combined, the expected number of light string queries until the simulation terminates, which is at most the expected number of queries required to reveal two $\mathrm{K}$-vertices, is bounded by $\frac{2n}{n_{\mathrm{K}}(s) - 2t}$.
    
 \end{proof}

\subsection{A hard to distinguish construction}

To construct a pair of hard-to-distinguish graphs, we define a few parameters. For given $\eps$, $n$ and $n_k$, let $d = n_k / 2n$ (so that $\sqrt{2nd} = n_k$). Based on $n$, $n_k$, we define $n_h$ and $n_\ell$ as following:
\begin{itemize}
    \item $n_h = \ceil{\sqrt{\eps} n^{1/4} d^{3/4}}$.
    \item $n_\ell = \ceil{\sqrt{\eps} n^{3/4} d^{1/4}}$ if $n_h \ge 2$, and otherwise $n_\ell = \ceil{\eps n d}$.
\end{itemize}

We state the following simple observation.
\begin{observation}
Let $n_\ell$ and $n_h$ be defined as above. Then,
\begin{enumerate}
    \item $n_h = 1$ if and only if $d \le 1 / (\eps^{2/3} n^{1/3})$.
    \item If $n_h = 1$, then $n_\ell + n_h \le 3n / \sqrt{nd}$.
    \item If $n_h \ge 2$, then $n_\ell + n_h \le 4 \sqrt{\eps}\, n^{3/4} d^{1/4}$.
\end{enumerate}
\end{observation}
\begin{proof}
The first statement is by definition.  For the second statement, if $n_h = 1$ then $n_\ell\le 2\eps n d$, and thus \[  n_\ell + n_h
        \le 2 \cdot \eps n d + 1
        \le 2 n / \sqrt{nd} + 1
        \le 3 n / \sqrt{nd}. \]
For the third statement, if $n_h \ge 2$ then
$$
n_\ell + n_h \le 2 \cdot \sqrt{\eps}\, n^{3/4} d^{1/4} + 2 \cdot \sqrt{\eps}\, n^{1/4} d^{3/4} \le 4 \sqrt{\eps}\, n^{3/4} d^{1/4}. 
$$
\end{proof}

We now define a distribution over hard-to-distinguish pairs of strings.

\begin{definition}[Hard-to-distinguish strings]
Let $n_k$, $n_h$, and $n_\ell$ be given. We draw a pair of strings $(\bs_1, \bs_2)$ as follows:
\begin{itemize}
    \item Choose a set $\bK \subseteq \{1,\ldots,n\}$ of size $n_k$ uniformly at random.
    \item Define $\bA_1 = \{1,\ldots,n\} \setminus \bK$.
    \item Define $\bs_1$ by
    $$
    \bs_1(i) =
    \begin{cases}
        \mathrm{K}, & i \in \bK,\\
        \mathrm{A}, & i \in \bA_1.
    \end{cases}
    $$
    \item Choose a set $\bH_2 \subseteq \bA_1$ of size $n_h$ uniformly at random.
    \item Choose a set $\bL_2 \subseteq (\bA_1 \setminus \bH_2)$ of size $n_\ell$ uniformly at random.
    \item Define $\bA_2 = \bA_1 \setminus (\bH_2 \cup \bL_2)$.
    \item Define $\bs_2$ by
    $$
    \bs_2(i) =
    \begin{cases}
        \mathrm{K}, & i \in \bK,\\
        \mathrm{H}, & i \in \bH_2,\\
        \mathrm{L}, & i \in \bL_2,\\
        \mathrm{A}, & i \in \bA_2.
    \end{cases}
    $$
\end{itemize}
\end{definition}

A pair of hard-to-distinguish graphs is obtained by sampling $(\bs_1, \bs_2)$ according to this distribution and defining the corresponding graphs $G_{\bs_1}$ and $G_{\bs_2}$.

\begin{lemma} \label{lm:pr-bound-explicit-sequence}
Let $(s_1, s_2)$ be a pair of strings drawn from the hard-to-distinguish distribution with parameters $0 < \eps \le \frac{1}{11}$, $n \ge 16$, and $1 \le n_k \le \frac{1}{2} n$.  
Let $I_1,\ldots,I_q$ be a sequence of graph queries, and let $(I_{t,1},\ldots,I_{t,M_t})_{t=1}^q$ be a fixed sequence of corresponding string-query translations. Then, the probability that the sequence of graph-query answers for $G_{s_1}$ differs from that for $G_{s_2}$ is bounded by
\[
2 \frac{n_h}{n} \sum_{t=1}^q M_t + 2 \cdot\frac{n_\ell + n_h}{n} \cdot q.
\]
\end{lemma}
\begin{proof}
    A local query (degree, neighbor) is translated into a single full-query ($M_t = 1$). For every such a query at index $i$, if $s_1(i) = \mathrm{K}$ then $s_2(i) = \mathrm{K}$ as well. Otherwise, if $s_1(i) = \mathrm{A}$, then the probability that $s_2(i) \ne \mathrm{A}$ is bounded by $\frac{n_\ell + n_h}{n - n_k}$. By Markov's inequality, the probability to differ is bounded by the expected number of differences, which is at most $\frac{n_\ell + n_h}{n - n_k} q \le 2 \frac{n_\ell + n_h}{n} q$.

    An independent-set query is translated into a sequence of light queries. For every such a query at index $i$, if $s_1(i) = \mathrm{K}$ then $s_2(i) = \mathrm{K}$ as well, and if $s_1(i) = \mathrm{A}$, then the probability that $s_2(i) = \mathrm{H}$ is bounded by $\frac{n_h}{n - n_k} \le \frac{2n_h}{n}$. The graph-query can differ only if one of the string queries reveals an $\mathrm{H}$-index. By Markov's inequality, this probability is bounded by the expected number of $\mathrm{H}$-occurrences, which is at most $\frac{2 n_h}{n} \sum_{t=1}^q M_t$.
\end{proof}

\begin{lemma} \label{lm:pr-bound-graph-distinguish-deterministic}
Let $0 < \eps \le \frac{1}{11}$, $n \ge 16$, $1 \le n_k \le \frac{1}{2} n$, and $R = \frac{1}{20} \min\left( n_k, \sqrt{\frac{n}{\eps n_k}} \right)$.  
Suppose that we draw a hard-to-distinguish pair $(\bs_1, \bs_2)$ based on the parameters $\eps$, $n$, and $n_k$.  Then, for any deterministic algorithm making at most $q$ graph queries in expectation, the probability that the sequences of query answers produced on inputs $G_{\bs_1}$ and $G_{\bs_2}$ differ is at most $q / R$.
\end{lemma}
\begin{proof}
Let $\bQ$ denote the number of graph queries performed by the algorithm on input
$G_{\bs_1}$, and  $q = \Ex[\bQ]$. Let $\bZ$ be the indicator random variable defined by
\[
\bZ \eqdef \indi\{\text{the executions on } G_{\bs_1}
\text{ and } G_{\bs_2} \text{ follow different execution paths}\}.
\]

We condition on the value of $\bQ$. If $\bQ = Q >  n_k/4$, then
\[
\Pr\left[\bZ = 1 \mid \bQ = Q\right] \le 1 \le Q/R,
\]
and the bound holds trivially. Hence, assume that $Q \le n_k/4$.

By Lemma~\ref{lm:lbnd:IS-translation-string-queries}, each graph query is translated into
a sequence of string queries whose expected length is at most $4n/n_k$. Let $\mathbf M_t$
denote the (random) length of the sequence corresponding to the $t$-th query, where the randomness is over the execution of the translation procedure. By linearity of
expectation and Lemma~\ref{lm:pr-bound-explicit-sequence}, when conditioned on $\bQ = Q$ for some $Q \le n_k / 4$,
\[
\Prx[\bZ = 1 \mid \bQ = Q]
\le
2\frac{n_h}{n}\sum_{t=1}^{Q}\Ex[\bM_t]
+
2\frac{n_\ell+n_h}{n}Q
\le
\left(8\frac{n_h}{n_k}+2\frac{n_\ell+n_h}{n}\right)Q.
\]

If $d \le 1/(\varepsilon^{2/3}n^{1/3})$, then $n_h=1$, and substituting the corresponding
bounds yields
\[
\Pr[\bZ = 1 \mid \bQ = Q] \le \frac{17}{n_k}Q \le Q/R.
\]
If $d > 1/(\varepsilon^{2/3}n^{1/3})$, then $n_h \ge 2$, and substituting the appropriate
bounds gives
\[
\Pr[\bZ = 1 \mid \bQ = Q]
\le
(20/2^{1/4})\sqrt{\varepsilon n_k/n}\cdot Q
\le
Q/R.
\]

Thus, for every realization of $\bQ$,
\[
\Pr[\bZ = 1 \mid \bQ = Q] \le Q/R.
\]
Applying the law of total probability,
\[
\Pr[\bZ = 1]
=
\Ex[\Pr[\bZ = 1 \mid \bQ]]
\le
\Ex[\bQ]/R
\le
q/R.
\]
\end{proof}

\begin{lemma} \label{lm:pr-bound-graph-distinguish}
    For given $0 < \eps \le \frac{1}{11}$, $n \ge 16$ and $1 \le n_k \le \frac{1}{2}n$, let  $R = \frac{1}{20} \min\left( n_k, \sqrt{\frac{n}{\eps n_k}} \right)$. Assume that we draw a hard-to-distinguish pair $(\bs_1, \bs_2)$ based on the parameters $\eps$, $n$ and $n_k$. 
    Then, for any randomized algorithm making at most $q$ graph queries in expectation, the probability that the sequences of query answers produced on inputs $G_{\bs_1}$ and $G_{\bs_2}$ differ is at most $q / R$.
\end{lemma}
\begin{proof}
  Applying Yao's principle~\cite{Yao77}, which states that the performance of any randomized algorithm can be analyzed as a distribution over deterministic algorithms, we immediately extend Lemma~\ref{lm:pr-bound-graph-distinguish-deterministic} to randomized algorithms. That is, for a randomized algorithm making at most $q$ graph queries in expectation, the probability that the query answer sequences for inputs $G_{\bs_1}$ and $G_{\bs_2}$ differ is still bounded by $q / R$.
\end{proof}

We can now finish with the proof of the lower bound.

\begin{proofof}{Theorem~\ref{th:lower-bound-main}}
    Consider a randomized algorithm $\calA$ that makes at most $q$ queries.

    We draw a $(\bs_1,\bs_2)$ according to the parameters $\eps' = 3\eps$, $n$ and $n_k = \floor{\sqrt{2m}}$. Note that for $m \ge 6$ we have that $\sqrt{m} \le n_k \le 2\sqrt{m}$.

    By Lemma~\ref{lm:pr-bound-graph-distinguish}, the query answer sequences of $\calA$ on $G_{\bs_1}$ and $G_{\bs_2}$ are $20 q / \min(n_k, \sqrt{n / (\eps n_k)} )$-close. Consequently, the probability that $\calA$ distinguishes $G_{\bs_1}$ from $G_{\bs_2}$ is bounded by
\begin{align*}
\Pr[\calA(G_{\bs_1}) \neq \calA(G_{\bs_2})] 
&\le \frac{1}{2} + \frac{20 q}{\min\left( n_k, \sqrt{\frac{n}{\eps n_k}} \right)} 
\le \frac{1}{2} + \frac{20 q}{\min\left( \sqrt{m}, \sqrt{\frac{n}{\eps \sqrt{2m}}} \right)} \\[1mm]
&< \frac{1}{2} + \frac{30 q}{\min\left( \sqrt{m}, \sqrt{\frac{n}{\eps \sqrt{m}}} \right)} 
= \frac{1}{2} + \frac{q}{R}.
\end{align*}

    Let $m_1$, $m_2$ be the number of edges in $G_{\bs_1}$ and $G_{\bs_2}$ respectively.
    Observe that $m_1 = \binom{n_k}{2} \le \frac{1}{2} (n_k)^2 \le m$ and that $m_2 \ge m_1 + \eps' m \ge (1 + \eps') m_1 = (1 + 3\eps) m_1$. If $\calA$ would estimate the number of edges in $G$ within $(1 \pm \eps)$-multiplicative error with probability at least $\frac{1}{2} + q/R$, then it would be a contradiction, since $(1 + \eps)m_1 < (1-\eps)(1+3\eps)m_1 \le (1 - \eps)m_2$.
\end{proofof}

\newpage

\phantomsection
\addcontentsline{toc}{section}{References}

\bibliographystyle{alpha}
\bibliography{main}

\appendix

\newcommand\procnameZextractZedge{\textsf{Extract-Edge}}
\newcommand\procnameZenumerateZedgesZbipartite{\textsf{Enumerate-Edges-Bipartite}}

\newpage
\section{$O(\log n)$-Amortized Edge Enumeration}
\label{apx:enumerate-edges-amortized}

In this appendix we prove Lemma~\ref{lm:enumerate-edges-amortized}. Throughout, we assume access to the graph via an independent-set oracle. We begin by recalling the edge-extraction primitive of~\cite{BHRRS18}.

\begin{lemma}[\cite{BHRRS18}, implicit] \label{lm:extract-edge}
    There exists a deterministic procedure $\procnameZextractZedge$ that takes as input a graph $G$ on $n$ vertices and a vertex set $A\subseteq [n]$, and either returns an arbitrary edge of $G$ with both endpoints in $A$, or correctly reports that $A$ is an independent set. The procedure performs at most $O(\log n)$ independent-set queries in the worst case.
\end{lemma}

Our proof is algorithmic: from the argument below, the reader can directly extract an explicit procedure achieving the desired guarantee. Given $A\subseteq [n]$ as an input, the algorithm proceeds in three phases:
\begin{enumerate}
    \item Constructing a vertex cover of the subgraph induced by $A$.
    \item Using this vertex cover to color the vertices of $A$.
    \item Enumerating all remaining edges using the induced coloring.
\end{enumerate}

\subsection{Enumerating Edges Between Two Independent Sets}

We first address the problem of enumerating edges between two independent vertex sets. This is achieved via a refinement of Beame's algorithm~\cite{BHRRS18}, with an explicit amortized bound.

\begin{lemma} \label{lm:enumerate-edges-bipartite-amortized}
    There exists a deterministic procedure $\procnameZenumerateZedgesZbipartite$ that, given a graph $G$ and two disjoint independent vertex sets $A$ and $B$, enumerates all edges with one endpoint in $A$ and the other in $B$. The $i$-th edge is reported after at most $O(1 + i \log n)$ independent-set queries. The procedure terminates after $O(1 + X \log n)$ queries, where $X$ is the total number of such edges.
\end{lemma}

\begin{proof}
The procedure follows the recursive partitioning approach of~\cite{BHRRS18}. We first issue a single independent-set query on $A \cup B$. If it is accepted, then there are no edges between $A$ and $B$, and the algorithm terminates after $O(1)$ queries.

Otherwise, we initialize a stack with the pair $(A,B)$. While the stack is nonempty, we pop a pair $(A_i,B_i)$ and proceed as follows:
\begin{itemize}
    \item If $|A_i| \ge 2$, partition $A_i$ into two subsets $A_{i,1}$ and $A_{i,2}$ of sizes $\lfloor |A_i|/2 \rfloor$ and $\lceil |A_i|/2 \rceil$. For each $r \in \{1,2\}$, if $A_{i,r} \cup B_i$ is not independent, push $(A_{i,r},B_i)$ onto the stack.
    \item Else, if $|A_i|=1$ and $|B_i| \ge 2$, symmetrically partition $B_i$ and recurse analogously.
    \item If $|A_i|=|B_i|=1$, then the unique pair $(u,v)$ with $u \in A_i$ and $v \in B_i$ is an edge, which we report.
\end{itemize}

Each pair pushed onto the stack is guaranteed to contain at least one edge. Consider the potential function
\[
\Phi(A_i,B_i) \eqdef \log_2 |A_i| + \log_2 |B_i|.
\]
If $\Phi(A_i,B_i) > 0$, then partitioning strictly decreases the potential by at least
\[
\min_{m \ge 2} \left(\log_2 m - \log_2 \ceil{m/2}\right) = \log_2(3/2) \ge \tfrac{1}{2}.
\]
If $\Phi(A_i,B_i)=0$, then an edge is reported, and the subsequent increase in potential is at most $2\log_2 n$ (since $\log_2 \abs{A_{i+1}} + \log_2 \abs{B_{i+1}} \le \log_2 n + \log_2 n$).

Since the initial potential is at most $2\log_2 n$, it follows that the first edge is reported after at most $O(\log n)$ queries, and each subsequent edge is reported after an additional $O(\log n)$ queries. This establishes the stated amortized bound.
\end{proof}

\subsection{Constructing a Vertex Cover}

We now show how to compute a vertex cover of the induced subgraph on $A$, while enumerating edges with amortized guarantees.

\begin{lemma} \label{lm:enumerate-edges-find-cover}
    There exists a deterministic algorithm that, given a graph $G$ and a vertex set $A$, outputs a list of pairwise disjoint edges whose endpoints form a vertex cover of the subgraph induced by $A$. The $i$-th edge is found after $O(1 + i \log n)$ independent-set queries.
\end{lemma}

\begin{proof}
The algorithm proceeds iteratively. In each iteration, we first query whether the subgraph induced by $A$ is independent. If it is, we terminate.

Otherwise, we invoke $\procnameZextractZedge$ to obtain an edge $uv$ with $u,v \in A$ using $O(\log n)$ queries. We output this edge and remove both endpoints from $A$.

The final iteration costs one query and outputs no edge; each earlier iteration outputs one edge at cost $O(\log n)$. This yields the stated amortized bound.

It remains to argue that the set of endpoint vertices of the found edges forms a vertex cover of the subgraph induced by the original vertex set. Let $E'$ be the set of edges reported by the algorithm, and let $V'$ be the set of their endpoints. Suppose, for the sake of contradiction, that there exists an edge $(x,y)$ in the subgraph induced by the original $A$ such that neither $x$ nor $y$ belongs to $V'$. Then both $x$ and $y$ were never removed from $A$, and hence were still present when the algorithm terminated. However, the algorithm terminates only when the remaining induced subgraph is independent, contradicting the existence of the edge $(x,y)$. Therefore, every edge in the induced subgraph has at least one endpoint in $V'$, and $V'$ is a vertex cover.
\end{proof}

\subsection{Enumerating Edges After Coloring}

We next handle the case where the vertex set has been partitioned into independent sets.

\begin{lemma} \label{lm:enumerate-edges-after-independent}
Let $k \ge 3$, and let $A_1,\ldots,A_k$ be disjoint independent vertex sets such that for every $1 \le i < j \le k-1$, the set $A_i \cup A_j$ is not independent. There exists a deterministic algorithm that enumerates all edges induced by $\bigcup_{i=1}^k A_i$. The $t$-th edge is reported after $O(t \log n)$ independent-set queries, and the algorithm terminates after $O(X \log n)$ queries, where $X$ is the total number of edges.
\end{lemma}

\begin{proof}
For each $1 \le i < j \le k$, let $X_{i,j}$ denote the number of edges between $A_i$ and $A_j$. We enumerate all pairs $(i,j)$ in lexicographic order by increasing $j$, breaking ties by increasing $i$, and invoke $\procnameZenumerateZedgesZbipartite$ on each pair.

By Lemma~\ref{lm:enumerate-edges-bipartite-amortized}, the number of queries performed before the $t$-th edge of $(A_i,A_j)$ is reported is bounded by
\[
\sum_{j'=1}^{j-1} \sum_{i'=1}^{j'-1} O(1 + X_{i',j'} \log n)
+ \sum_{i'=1}^{j-1} O(1 + X_{i',j} \log n)
+ O(1 + t \log n).
\]

Since $X_{i,j} \ge 1$ for all $1 \le i < j \le k-1$, the above expression simplifies to:
\[
\sum_{j'=1}^{j-1} \sum_{i'=1}^{j'-1} O(X_{i',j'} \log n)
+ \sum_{i'=1}^{j-1} O(1 + X_{i',j} \log n)
+ O(1 + t \log n).
\]

Which can be represented as:
\[
O\!\left(j + t' \log n\right),\qquad t' = \sum_{j'=1}^{j-1} \sum_{i'=1}^{j'-1} X_{i',j'} +  \sum_{i'=1}^{j-1} X_{i',j} + t,
\]
where $t'$ is the global index of the reported edge in the overall enumeration. Since at least $j-2$ edges must have been reported before reaching any edge incident to $A_j$, we have $t' \ge (j-2)+1 \ge j/2$ (for $j \ge 2$). Thus $O(j + t' \log n)=O(t' \log n)$, as claimed.

For termination, observe that terminating (by figuring out that there are no additional edges) costs $O(k)$ queries (in the worst case where $A_k$ has no edges to $\bigcup_{i=1}^{k-1} A_i$), which is dominated by $O(X \log n)$ (since $X \ge \binom{k-1}{2} = \Omega(k^2)$).
\end{proof}

\subsection{Proof of the Main Amortized Enumeration Lemma}

We now prove the amortized version of Lemma \ref{lm:enumerate-edges-amortized-formal}.

\begin{lemma}[Identical to Lemma \ref{lm:enumerate-edges-amortized}] \label{lm:enumerate-edges-amortized-formal}
    There exists an implementation of $\procnameZenumerateZedges$ such that the $t$-th edge is reported after $O(1 + t \log n)$ independent-set queries. The algorithm terminates after $O(1 + X \log n)$ queries, where $X$ is the total number of edges.
\end{lemma}

\begin{proof}
We first apply Lemma~\ref{lm:enumerate-edges-find-cover} to the input vertex set $A$, obtaining a collection of $X_1$ disjoint edges at a total cost of $O(1 + X_1 \log n)$ queries. Let $V_1$ denote the set of vertices incident to these edges.

Next, we greedily color the vertices in $V_1$. We maintain independent sets $A_1,A_2,\ldots$, initially empty. Each vertex $u \in V_1$ is assigned to the smallest index $i$ such that $A_i \cup \{u\}$ remains independent. During this process, whenever an independent-set query rejects, we invoke $\procnameZextractZedge$ to obtain an edge incident to $u$, at cost $O(\log n)$.

If $u$ is assigned to $A_{i_u}$, then its coloring costs $O(1 + (i_u-1)\log n)$ queries and yields exactly $i_u-1$ discovered edges (the $X_1$ edges found in the first phase are ``discovered'' for the second time, but not more than that).

Let $k$ be the smallest index such that $A_k$ is empty, and redefine $A_k = A \setminus V_1$. Note that $A \setminus V_1$ is independent, since $V_1$ is a vertex cover of the subgraph induced by $A$'s vertices. Also, $k \ge 3$ unless the graph is empty, since the very first edge must contribute vertices to $A_1$ and $A_2$. The sequence $A_1,\ldots,A_k$ satisfies the conditions of Lemma \ref{lm:enumerate-edges-after-independent}, which we apply to enumerate all remaining edges.

Each edge may be discovered at most once per phase, which is at most three times. We can ``charge'' every edge for three encounters once it is discovered for the first time, and ``consume'' the extra charge on further encounters. Since each phase contributes at most $O(\log n)$ amortized cost per edge, the $t$-th edge is reported after $O(1 + t \log n)$ independent-set queries overall.
\end{proof}

\end{document}